\documentclass[letterpaper, 10 pt, conference]{ieeeconf}
\IEEEoverridecommandlockouts
\usepackage[T1]{fontenc}
\usepackage{graphicx}
\usepackage{amssymb}
\usepackage{mathtools}
\usepackage{url}
\usepackage{microtype}
\usepackage{tikz}
\usetikzlibrary{positioning, arrows.meta, angles, quotes}
\usepackage{flushend}
\let\labelindent\relax
\usepackage{enumitem}

\newtheorem{remark}{\normalfont Remark}

\title{\LARGE \bf
Safety-Critical Formation Control of Non-Holonomic Multi-Robot Systems in Communication-Limited Environments
}

\author{Vishrut Bohara and Siavash Farzan%
\thanks{The authors are with the Electrical Engineering Department, California Polytechnic State University, San Luis Obispo, CA 93407, USA (corresponding e-mail: sfarzan@calpoly.edu)}}

\begin{document}

\maketitle

\begin{abstract}
This paper proposes a decentralized estimator-based safety-critical controller for formation control of non-holonomic multi-robot systems operating without inter-agent communication. Each follower reconstructs its predecessor's motion with a closed-form estimator and applies closed-form control barrier function (CBF) based laws that are robust to the resulting estimation error, enforcing collision avoidance and prescribed longitudinal and lateral formation offsets without online optimization. Lyapunov analysis establishes exponential convergence of the estimator under constant predecessor velocities and input-to-state-type ultimate boundedness under time-varying motion, from which the estimation-error bounds used by the robust CBF constraints are derived. The longitudinal spacing policy is shown to be $\mathcal{L}_2$ string stable under an explicit gain condition. Simulations and Gazebo experiments demonstrate accurate formation tracking, safety preservation, and disturbance attenuation.
\end{abstract}

\section{Introduction}

Multi-robot formations are important in warehousing, inspection, cooperative transport, and field robotics, where teams must maintain prescribed spatial relationships while remaining safe under sensing uncertainty and limited communication~\cite{review2017,2015survey}. For non-holonomic platforms, this problem is especially challenging when inter-agent links are unavailable or unreliable, since followers must infer predecessor motion from local measurements while reacting fast enough to preserve formation integrity and avoid collisions.

Observer-based formation controllers remove the need to communicate neighbor velocities by reconstructing them from relative measurements~\cite{quadrotor2023,aditya2019formation,moorthy2023formation}, and bearing- or orientation-based schemes relax the sensing requirements further~\cite{Bearing2024,orientation2023}; however, these works do not address inter-agent safety. Conversely, CBF-based safety filters provide formal collision-avoidance guarantees for multi-robot teams~\cite{ames2019control,wang2017safety}, and recent formation-control results combine them with tracking objectives~\cite{butler2024collaborative,rai2024safe,li2024safe,suganda2024perception,ballotta2024safe}. These methods, however, share one or more of the following limitations: (i) neighbors' velocities or intended actions are assumed to be communicated~\cite{butler2024collaborative,suganda2024perception,ballotta2024safe}; (ii) a quadratic program (QP) is solved at every control step~\cite{butler2024collaborative,rai2024safe,suganda2024perception}, whose feasibility and latency are difficult to certify on embedded hardware; (iii) agents are modeled as integrators or point masses~\cite{rai2024safe,li2024safe,wang2017safety} rather than non-holonomic vehicles; and (iv) the propagation of leader disturbances through the formation is not analyzed. In our earlier work~\cite{TVT2024}, estimation-based closed-form CBF control was developed for one-dimensional vehicular platoons; the planar non-holonomic setting considered here introduces follower rotation, lateral coupling between the two channels, and multi-branch formation graphs, none of which are covered by that analysis. The gap addressed here is thus a controller for non-holonomic robots that is simultaneously communication-free, closed-form, safe with respect to the estimation error it induces, and string stable.\looseness=-1

To this end, each follower reconstructs predecessor motion with a Lyapunov-based estimator and applies analytic CBF-based control laws that enforce safe longitudinal and lateral spacing. The resulting controller has deterministic computation, making it suitable for embedded and timing-sensitive robotic platforms. The main contributions are:
\begin{enumerate}[leftmargin=12pt]
    \item[i.] A closed-form estimator of non-holonomic predecessor motion that converges exponentially under constant predecessor velocity for arbitrary follower rotation, and is uniformly ultimately bounded, with an explicit bound, under time-varying motion.
    \item[ii.] A closed-form decentralized safety-critical controller whose robust CBF margins are tied to the estimator bound, with proofs of forward invariance and formation tracking without communication or online optimization, and a discussion of input saturation.
    \item[iii.] An explicit sufficient condition for $\mathcal{L}_2$ string stability of the longitudinal channel, and validation in simulation and Gazebo.
\end{enumerate}

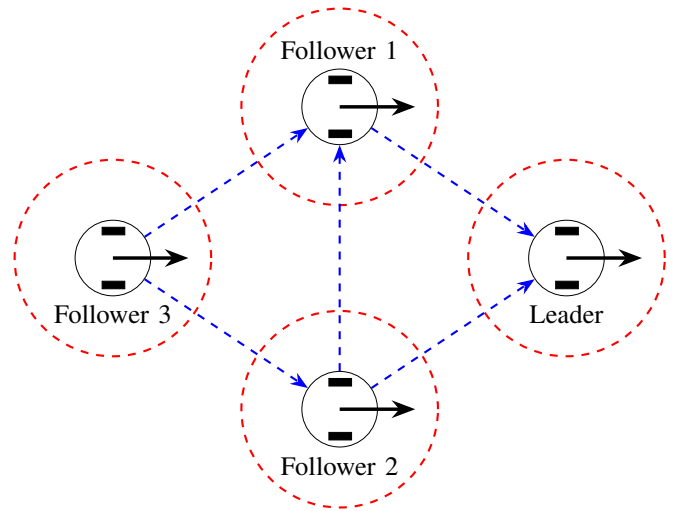
\begin{figure}[t]
    \centering
    \tikzset{every picture/.style={line width=0.25pt}}

\usetikzlibrary{calc}

\begin{tikzpicture}[>=Stealth, node distance=2cm,scale=0.75]

    \tikzstyle{wheel}=[rectangle, draw, fill=black, minimum width=0.3cm, minimum height=0.1cm]
    \tikzstyle{safety}=[red, thick, dashed]
    \tikzstyle{control}=[blue, thick, dashed, ->]
    \node[circle, draw, minimum size=1cm, label=below:{\scriptsize Leader}] (A1) at (0,0) {};
    \node[circle, draw, minimum size=1cm, label=above:{\scriptsize Follower 1}] (A2) at (-3,2) {};
    \node[circle, draw, minimum size=1cm, label=below:{\scriptsize Follower 2}] (A3) at (-3,-2) {};
    \node[circle, draw, minimum size=1cm, label=below:{\scriptsize Follower 3}] (A4) at (-6,0) {};

    \foreach \agent in {A1, A2, A3, A4} {
        \draw[wheel] ($(\agent.south west)!0.3!(\agent.south east)$) ++(0,-0.05) rectangle ++(0.3,0.1);
        \draw[wheel] ($(\agent.north west)!0.3!(\agent.north east)$) ++(0,-0.05) rectangle ++(0.3,0.1);
    }

    \draw[red,thick,dashed] (0,0) circle (1.3cm);
    \draw[red,thick,dashed] (-3,2) circle (1.3cm);
    \draw[red,thick,dashed] (-3,-2) circle (1.3cm);
    \draw[red,thick,dashed] (-6,0) circle (1.3cm);

    \draw[->,color={rgb, 255:red, 0; green, 0; blue, 255 }  ,draw opacity=1 , thick, dashed] (A2) -- (A1);
    \draw[->,color={rgb, 255:red, 0; green, 0; blue, 255 }  ,draw opacity=1 , thick, dashed] (A3) -- (A1);
    \draw[->,color={rgb, 255:red, 0; green, 0; blue, 255 }  ,draw opacity=1 , thick, dashed] (A3) -- (A2);
    \draw[->,color={rgb, 255:red, 0; green, 0; blue, 255 }  ,draw opacity=1 , thick, dashed] (A4) -- (A3);
    \draw[->,color={rgb, 255:red, 0; green, 0; blue, 255 }  ,draw opacity=1 , thick, dashed] (A4) -- (A2);

    \draw[->, line width=1.3pt] (A1.center) -- +(0:1.0) node[above right] {};
    \draw[->, line width=1.3pt] (A2.center) -- +(0:1.0) node[above right] {};
    \draw[->, line width=1.3pt] (A3.center) -- +(0:1.0) node[above right] {};
    \draw[->, line width=1.3pt] (A4.center) -- +(0:1.0) node[above right] {};

\end{tikzpicture}
    \vspace{-25pt}
    \caption{Leader-Follower formation control of mobile robots in a diamond formation, illustrating inter-agent interaction edges and safety boundaries. The red dashed circles denote the safety regions for each robot, and the blue dashed arrows represent predecessor-follower interactions, where followers maintain desired distances from their predecessors without communication.}
    \label{fig:setup_2d}
\end{figure}

\section{Problem Description} \label{sec:problem_description}

We consider decentralized leader-follower formation control for non-holonomic robots with no inter-agent communication. Each follower measures its own speed and the relative pose of relevant predecessors, and must maintain prescribed longitudinal and lateral offsets while guaranteeing collision avoidance. The leader’s center of curvature is assumed to lie outside the formation so that the desired geometry is feasible.

\vspace{-5pt}
\subsection{Formation Control}\label{sec:formation_control}
\begin{figure}[t]
    \centering
    \tikzset{every picture/.style={line width=0.75pt}}

\begin{tikzpicture}[>=Stealth]

    \draw[dashed,->] (0,0) -- (6,0) node[right] {$x$};
    \draw[dashed,->] (0,0) -- (0,2.5) node[above] {$y$};
    \draw[dashed,->] (4,2) -- (5.5,2) node[right] {};
    
    \node[circle, draw, minimum size=1.6cm, line width=1.5pt, label=below:Follower] (A1) at (0,0) {};
    \draw[->, line width=1.3pt] (A1.center) -- +(1.8,0) node[below,yshift=-3pt] {$v, u$};
    
    \node[circle, draw, minimum size=1.6cm, line width=1.5pt, label=above:Predecessor] (A2) at (4,2) {};
    \draw[->, line width=1.3pt] (A2.center) -- +(30:1.8) node[above right,yshift=-5pt] {$v^{p}, u^{p}$};
    
    \draw[dashed, ->] (A1.center) -- node[midway, below] {$d$} (A2.center);
    
    \draw[dashed] (A1.center) -- node[midway, above] {$d_x$} (A2.center |- A1.center);
    \draw[dashed] (A2.center) -- node[midway, left] {$d_y$} (A2.center |- A1.center);

    \coordinate (a) at (2,0);
    \coordinate (a1) at (2,-2);
    \coordinate (b) at (7,2);
    \coordinate (c) at (7,3.8);

    \pic [draw, ->, "$\psi$", angle eccentricity=1.3,angle radius=1.2cm] {angle = b--A2--c};

    \pic [draw, ->, "$\theta$", angle eccentricity=1.3,angle radius=1.2cm] {angle=a--A1--A2};
    \pic [draw, ->, "$\omega$", angle eccentricity=1.8,angle radius=0.3cm] {angle = a--A1--a1};
    \pic [draw, ->, angle eccentricity=1.1,angle radius=0.3cm] {angle = b--A2--a1};
    \node at (3.5,2.2) {$\omega^p$};
\end{tikzpicture}
    \vspace{-25pt}
    \caption{Predecessor-follower pair and associated parameters in a leader-follower formation control setting.}
    \label{fig:pred-suc-FC}
\end{figure}
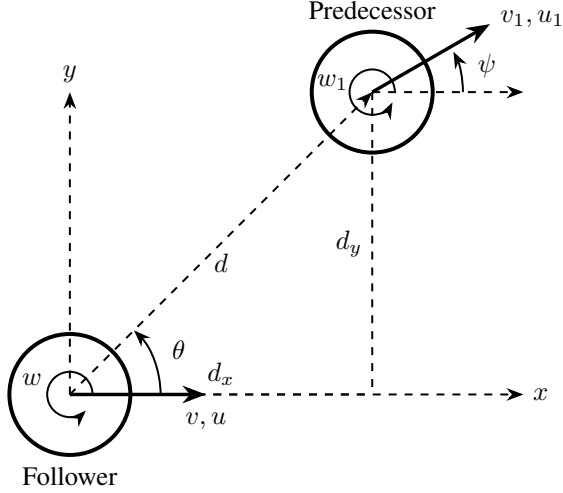

The formation is modeled as a directed acyclic graph \(G=(V,E)\), where
\(
V=\{L,F_1,F_2,\ldots,F_n\},
\)
with \(L\) the leader and \(F_i\) the followers. Each follower regulates a longitudinal offset \(d_x^*\) relative to its \(X^+\)-edge predecessor and a signed lateral offset \(d_y^*\) relative to its \(Y\)-edge predecessor. The scalar \(d_s>0\) denotes the CBF safety margin and is distinct from the tracking setpoints. Figs.~\ref{fig:setup_2d} and~\ref{fig:pred-suc-FC}  illustrate the formation topology and the predecessor-follower geometry used in the controller design, respectively.

\vspace{-5pt}
\subsection{Predecessor-Follower Dynamics}
We now consider a generic predecessor-follower pair. Let the \(i^{\text{th}}\) robot track its predecessor, the \((i-1)^{\text{th}}\) robot. In the follower frame, \(d\) denotes the inter-agent distance, \(\theta\) the bearing angle, \(v^{p}\) and \(v\) the predecessor and follower linear velocities, and \(\psi\) the relative heading. The follower measures \(d\), \(\theta\), and \(v\), and computes the control inputs \(u\) and \(\omega\). The pair dynamics are
\begin{equation}
    \begin{aligned}
        \dot{d}      &= v^{p} \cos(\theta - \psi) - v \cos\theta, \\
        \dot{\theta} &= \frac{v \sin\theta - v^{p} \sin(\theta - \psi)}{d} - \omega, \\
        \dot{v}^{p}  &= u^{p}, \qquad \dot{v} = u, \qquad
        \dot{\psi}   = \omega^{p} - \omega.
    \end{aligned}\label{dyn_2d:1}
\end{equation}

For analysis and control design, we project the dynamics onto the follower frame using
\[
d_x=d\cos\theta,\quad d_y=d\sin\theta,
\]
\[
v_x^{p}=v^{p}\cos\psi,\quad v_y^{p}=v^{p}\sin\psi,\]
\[
a_x^{p}=u^{p}\cos\psi-v_y^{p}\omega^{p},\quad
a_y^{p}=u^{p}\sin\psi+v_x^{p}\omega^{p}.
\]
This yields
\begin{equation}
    \begin{aligned}
        \dot{d}_x   &= v_x^{p} - v + d_y\,\omega, \qquad
        &\dot{d}_y   &= v_y^{p} - d_x\,\omega, \\
        \dot{v}_x^{p} &= a_x^{p} + v_y^{p}\,\omega,
        &\dot{v}_y^{p} &= a_y^{p} - v_x^{p}\,\omega.
    \end{aligned}\label{dyn_2d:2}
\end{equation}

The admissible inputs satisfy
\begin{equation}
    \begin{aligned}
        &0 \leq v \leq v_{\max}, \quad |u| \leq u_{\max}, \quad |\omega| \leq \omega_{\max},\\[0pt]
        &|a_x^{p}|,\,|a_y^{p}| \leq u_{\max} + v_{\max}\,\omega_{\max} \triangleq a_{\max}.
    \end{aligned}\label{bnd:1}
\end{equation}

\subsection{String Stability}
Following~\cite{orosz16,swaioa94}, a predecessor-follower chain is $\mathcal{L}_2$ string stable if the gain $\mathcal{S}$ from a predecessor's velocity perturbation to the follower's velocity perturbation satisfies \(\mathcal{S}\le 1\) along every edge, so that disturbances do not amplify as they propagate from the leader.
In addition, safe operation requires the inter-agent distances to remain above collision-avoidance thresholds while staying close to the desired offsets. The controller in Section~\ref{sec:methodology} enforces these objectives through Lyapunov and CBF constraints.

\section{Estimator-based Safety Critical Controller}\label{sec:methodology}

\begin{figure}[t]
    \centering
    \tikzset{every picture/.style={line width=0.75pt}}   

\begin{tikzpicture}[x=0.75pt,y=0.75pt,yscale=-0.9,xscale=0.9]

\draw  [color={rgb, 255:red, 74; green, 74; blue, 74 }  ,draw opacity=1 ][fill={rgb, 255:red, 0; green, 0; blue, 255 }  ,fill opacity=0.15 ] (230,78) .. controls (230,73.58) and (233.58,70) .. (238,70) -- (332,70) .. controls (336.42,70) and (340,73.58) .. (340,78) -- (340,102) .. controls (340,106.42) and (336.42,110) .. (332,110) -- (238,110) .. controls (233.58,110) and (230,106.42) .. (230,102) -- cycle ;
\draw  [color={rgb, 255:red, 74; green, 74; blue, 74 }  ,draw opacity=1 ][fill={rgb, 255:red, 155; green, 155; blue, 155 }  ,fill opacity=0.3 ] (10,123) .. controls (10,118.58) and (13.58,115) .. (18,115) -- (112,115) .. controls (116.42,115) and (120,118.58) .. (120,123) -- (120,147) .. controls (120,151.42) and (116.42,155) .. (112,155) -- (18,155) .. controls (13.58,155) and (10,151.42) .. (10,147) -- cycle ;
\draw  [color={rgb, 255:red, 74; green, 74; blue, 74 }  ,draw opacity=1 ][fill={rgb, 255:red, 0; green, 0; blue, 255 }  ,fill opacity=0.15 ] (75,168) .. controls (75,163.58) and (78.58,160) .. (83,160) -- (177,160) .. controls (181.42,160) and (185,163.58) .. (185,168) -- (185,192) .. controls (185,196.42) and (181.42,200) .. (177,200) -- (83,200) .. controls (78.58,200) and (75,196.42) .. (75,192) -- cycle ;
\draw    (185,170) -- (227,170) ;
\draw [shift={(230,170)}, rotate = 180] [fill={rgb, 255:red, 0; green, 0; blue, 0 }  ][line width=0.08]  [draw opacity=0] (5.36,-2.57) -- (0,0) -- (5.36,2.57) -- cycle    ;
\draw    (205,180) -- (227,180) ;
\draw [shift={(230,180)}, rotate = 180] [fill={rgb, 255:red, 0; green, 0; blue, 0 }  ][line width=0.08]  [draw opacity=0] (5.36,-2.57) -- (0,0) -- (5.36,2.57) -- cycle    ;
\draw    (205,105) -- (227,105) ;
\draw [shift={(230,105)}, rotate = 180] [fill={rgb, 255:red, 0; green, 0; blue, 0 }  ][line width=0.08]  [draw opacity=0] (5.36,-2.57) -- (0,0) -- (5.36,2.57) -- cycle    ;
\draw    (205,105) -- (205,135) ;
\draw    (120,135) -- (205,135) ;
\draw    (130,225) -- (130,203) ;
\draw [shift={(130,200)}, rotate = 90] [fill={rgb, 255:red, 0; green, 0; blue, 0 }  ][line width=0.08]  [draw opacity=0] (5.36,-2.57) -- (0,0) -- (5.36,2.57) -- cycle    ;
\draw    (130,113) -- (130,157) ;
\draw [shift={(130,160)}, rotate = 270] [fill={rgb, 255:red, 0; green, 0; blue, 0 }  ][line width=0.08]  [draw opacity=0] (5.36,-2.57) -- (0,0) -- (5.36,2.57) -- cycle    ;
\draw [shift={(130,110)}, rotate = 90] [fill={rgb, 255:red, 0; green, 0; blue, 0 }  ][line width=0.08]  [draw opacity=0] (5.36,-2.57) -- (0,0) -- (5.36,2.57) -- cycle    ;
\draw  [color={rgb, 255:red, 74; green, 74; blue, 74 }  ,draw opacity=1 ][fill={rgb, 255:red, 0; green, 0; blue, 255 }  ,fill opacity=0.15 ] (230,168) .. controls (230,163.58) and (233.58,160) .. (238,160) -- (332,160) .. controls (336.42,160) and (340,163.58) .. (340,168) -- (340,192) .. controls (340,196.42) and (336.42,200) .. (332,200) -- (238,200) .. controls (233.58,200) and (230,196.42) .. (230,192) -- cycle ;
\draw  [color={rgb, 255:red, 74; green, 74; blue, 74 }  ,draw opacity=1 ][fill={rgb, 255:red, 0; green, 0; blue, 255 }  ,fill opacity=0.15 ] (75,78) .. controls (75,73.58) and (78.58,70) .. (83,70) -- (177,70) .. controls (181.42,70) and (185,73.58) .. (185,78) -- (185,102) .. controls (185,106.42) and (181.42,110) .. (177,110) -- (83,110) .. controls (78.58,110) and (75,106.42) .. (75,102) -- cycle ;
\draw  [color={rgb, 255:red, 74; green, 74; blue, 74 }  ,draw opacity=1 ][fill={rgb, 255:red, 155; green, 155; blue, 155 }  ,fill opacity=0.3 ] (10,230) .. controls (10,227.24) and (12.24,225) .. (15,225) -- (240,225) .. controls (242.76,225) and (245,227.24) .. (245,230) -- (245,245) .. controls (245,247.76) and (242.76,250) .. (240,250) -- (15,250) .. controls (12.24,250) and (10,247.76) .. (10,245) -- cycle ;
\draw    (130,215) -- (205,215) ;
\draw    (205,180) -- (205,215) ;
\draw    (340,180) -- (357,180) ;
\draw [shift={(360,180)}, rotate = 180] [fill={rgb, 255:red, 0; green, 0; blue, 0 }  ][line width=0.08]  [draw opacity=0] (5.36,-2.57) -- (0,0) -- (5.36,2.57) -- cycle    ;
\draw    (130,45) -- (130,67) ;
\draw [shift={(130,70)}, rotate = 270] [fill={rgb, 255:red, 0; green, 0; blue, 0 }  ][line width=0.08]  [draw opacity=0] (5.36,-2.57) -- (0,0) -- (5.36,2.57) -- cycle    ;
\draw  [color={rgb, 255:red, 74; green, 74; blue, 74 }  ,draw opacity=1 ][fill={rgb, 255:red, 155; green, 155; blue, 155 }  ,fill opacity=0.3 ] (10,25) .. controls (10,22.24) and (12.24,20) .. (15,20) -- (240,20) .. controls (242.76,20) and (245,22.24) .. (245,25) -- (245,40) .. controls (245,42.76) and (242.76,45) .. (240,45) -- (15,45) .. controls (12.24,45) and (10,42.76) .. (10,40) -- cycle ;
\draw    (185,90) -- (227,90) ;
\draw [shift={(230,90)}, rotate = 180] [fill={rgb, 255:red, 0; green, 0; blue, 0 }  ][line width=0.08]  [draw opacity=0] (5.36,-2.57) -- (0,0) -- (5.36,2.57) -- cycle    ;
\draw    (130,55) -- (200,55) ;
\draw    (200,55) -- (200,75) ;
\draw    (200,75) -- (227,75) ;
\draw [shift={(230,75)}, rotate = 180] [fill={rgb, 255:red, 0; green, 0; blue, 0 }  ][line width=0.08]  [draw opacity=0] (5.36,-2.57) -- (0,0) -- (5.36,2.57) -- cycle    ;
\draw    (340,90) -- (357,90) ;
\draw [shift={(360,90)}, rotate = 180] [fill={rgb, 255:red, 0; green, 0; blue, 0 }  ][line width=0.08]  [draw opacity=0] (5.36,-2.57) -- (0,0) -- (5.36,2.57) -- cycle    ;

\draw (9,121) node [anchor=north west][inner sep=0.75pt]  [font=\small] [align=left] {\begin{minipage}[lt]{73.64pt}\setlength\topsep{0pt}
\begin{center}
Measurements of\\internal state
\end{center}

\end{minipage}};
\draw (4,22) node [anchor=north west][inner sep=0.75pt]  [font=\small] [align=left] {\begin{minipage}[lt]{166.78pt}\setlength\topsep{0pt}
\begin{center}
Measurements for $X^+$-edge predecessor
\end{center}

\end{minipage}};
\draw (72,166) node [anchor=north west][inner sep=0.75pt]  [font=\small] [align=left] {\begin{minipage}[lt]{76.21pt}\setlength\topsep{0pt}
\begin{center}
Estimator for $Y$\\edge predecessor
\end{center}

\end{minipage}};
\draw (73,74) node [anchor=north west][inner sep=0.75pt]  [font=\small] [align=left] {\begin{minipage}[lt]{76.21pt}\setlength\topsep{0pt}
\begin{center}
Estimator for $X^+$\\edge predecessor
\end{center}

\end{minipage}};
\draw (228,74) node [anchor=north west][inner sep=0.75pt]  [font=\small] [align=left] {\begin{minipage}[lt]{75.17pt}\setlength\topsep{0pt}
\begin{center}
$X^+$-edge\\tracking controller
\end{center}

\end{minipage}};
\draw (6,230) node [anchor=north west][inner sep=0.75pt]  [font=\small] [align=left] {\begin{minipage}[lt]{161.09pt}\setlength\topsep{0pt}
\begin{center}
Measurements for $Y$-edge predecessor
\end{center}

\end{minipage}};
\draw (91,205) node [anchor=north west][inner sep=0.75pt]  [font=\small]  {$d_{x} ,d_{y}$};
\draw (133,120) node [anchor=north west][inner sep=0.75pt]  [font=\small]  {$v,\omega$};
\draw (187,73) node [anchor=north west][inner sep=0.75pt]  [font=\footnotesize]  {$\hat{d}_{x},\hat{v}_x^{p}$};
\draw (344,165) node [anchor=north west][inner sep=0.75pt]  [font=\small]  {$\omega$};
\draw (228,165) node [anchor=north west][inner sep=0.75pt]  [font=\small] [align=left] {\begin{minipage}[lt]{75.17pt}\setlength\topsep{0pt}
\begin{center}
$Y$-edge\\tracking controller
\end{center}

\end{minipage}};
\draw (187,148) node [anchor=north west][inner sep=0.75pt]  [font=\small]  {$\hat{d}_{y} ,\hat{v}_y^{p}$};
\draw (344,73) node [anchor=north west][inner sep=0.75pt]  [font=\small]  {$u$};
\draw (91,50) node [anchor=north west][inner sep=0.75pt]  [font=\small]  {$d_{x} ,d_{y}$};

\end{tikzpicture}
    \vspace{-24pt}
    \caption{Block diagram of the proposed estimator-integrated safety-critical control system, illustrating the interconnection of the estimator and controller modules within the decentralized formation control framework.}
    \label{fig:block-diagram}
\end{figure}
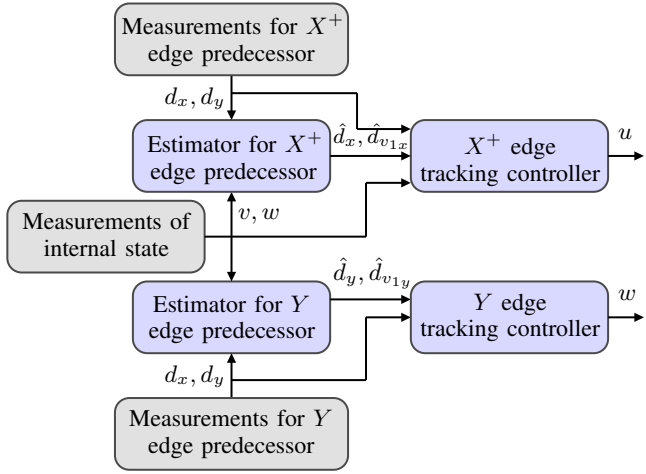

In the absence of explicit communication, each agent must estimate the velocity components of its predecessor to achieve robust formation tracking. Fig.~\ref{fig:block-diagram} provides a schematic of the estimator-based safety-critical control architecture, which integrates sensing (distance and angular measurements), computation (state estimation and control synthesis), and actuation in a distributed loop. Its closed-form structure yields deterministic runtimes suitable for embedded and resource-constrained platforms.

\vspace{-5pt}
\subsection{Estimator Design}\label{subsec:estimator_design}
Define the measurable quantities in the follower’s  frame:
\[ d_x = d \cos\theta,\quad d_y = d \sin\theta, \]
and let the true state vector for the predecessor be
\[ x = \begin{bmatrix} d_x,\; v_x^{p},\; d_y,\; v_y^{p} \end{bmatrix}^T. \]
The estimator generates the corresponding state estimates
\[ \hat{x} = \begin{bmatrix} \hat{d}_x,\; \hat{v}_x^{p},\; \hat{d}_y,\; \hat{v}_y^{p} \end{bmatrix}^T, \]

Let $g_d<0$, $g_v<0$, and $p$ be design gains, with $p$ tied to $g_d,g_v$ through $r^2 + r\,g_d - g_v = 0$, $r>1$, and $p = -\tfrac{r}{2}$ (Theorem~1). Assuming the follower has access to its own linear velocity $v$, angular velocity $\omega$, and the relative measurements $d$ and $\theta$, the proposed estimator is given by
\begin{equation}
    \begin{aligned}
        \dot{\hat{d}}_x &= \hat{v}_x^{p} - v + d\,\omega\,\sin\theta + g_d\,\tilde{d}_x,\\
        \dot{\hat v}_x^{p} &= g_v\,\tilde{d}_x + \hat{v}_y^{p}\,\omega + p\,\omega\,\tilde{d}_y,\\
        \dot{\hat{d}}_y &= \hat{v}_y^{p} - d\,\omega\,\cos\theta + g_d\,\tilde{d}_y,\\
        \dot{\hat v}_y^{p} &= g_v\,\tilde{d}_y - \hat{v}_x^{p}\,\omega - p\,\omega\,\tilde{d}_x.
    \end{aligned}\label{est_2d:1}
\end{equation}
where the estimation errors are defined as
\[ \tilde{e} = \hat{x} - x = \begin{bmatrix} \tilde{d}_x,\; \tilde{v}_x^{p},\; \tilde{d}_y,\; \tilde{v}_y^{p} \end{bmatrix}^T. \]

Using the estimates, the predecessor’s heading and speed can be reconstructed as
\begin{equation}
    \hat{\psi} = \arctan\!\left(\frac{\hat{v}_y^{p}}{\hat{v}_x^{p}}\right), \qquad
    \hat{v}^{p} = \sqrt{{\hat{v}_x^{p}}{^2} + {\hat{v}_y^{p}}{^2}}.
\end{equation}

The error dynamics, obtained by subtracting (\ref{dyn_2d:2}) from (\ref{est_2d:1}), can be expressed in compact form:
\begin{equation}
    \dot{\tilde{e}} =
    \underbrace{
    \begin{bmatrix}
        g_d & 1 & 0 & 0\\[0pt]
        g_v & 0 & p\,\omega & \omega\\[0pt]
        0 & 0 & g_d & 1\\[0pt]
        -p\,\omega & -\omega & g_v & 0
    \end{bmatrix}
    }_{A(\omega)}
    \tilde{e} \;+\;
    \underbrace{
    \begin{bmatrix}
        0 & 0\\[0pt]
        -1 & 0\\[0pt]
        0 & 0\\[0pt]
        0 & -1
    \end{bmatrix}
    }_{B}
    \begin{bmatrix}
        a_x^{p} \\
        a_y^{p}
    \end{bmatrix}.
    \label{est_2d:3}
\end{equation}
Here, the matrix $A(\omega)$ explicitly depends on the angular velocity $\omega$, and the term involving $a_x^{p}$ and $a_y^{p}$ accounts for the influence of the predecessor’s acceleration in the follower’s frame.\looseness=-1

\begin{remark}\label{rem:why_estimate}
Although $d_x$ and $d_y$ are measured, they are retained as estimator states so that $\hat v^{p}_x,\hat v^{p}_y$ are obtained through the output-injection terms $g_d\tilde d,\,g_v\tilde d$ rather than by differentiating $d_x,d_y$: each channel of \eqref{est_2d:1} is a second-order Luenberger-type observer whose bandwidth is set by $g_d,g_v$, so measurement noise is low-pass filtered rather than amplified. The position-error states are also what the $\omega$-coupling terms with gain $p$ act on, and these terms are what make Theorem~1 hold for arbitrary $\omega(t)$.
\end{remark}

\subsection{Estimator Stability}\label{subsec:stability}

Let $s_x \triangleq r\tilde d_x - \tilde v_x^{p}$, $s_y \triangleq r\tilde d_y - \tilde v_y^{p}$, and define
\begin{align}
    V_1 &= \tfrac{1}{2}\big(s_x^2 + |g_v|\,\tilde{d}_x^2 + \tilde v_x^{p\,2}\big), \nonumber\\
    V_2 &= \tfrac{1}{2}\big(s_y^2 + |g_v|\,\tilde{d}_y^2 + \tilde v_y^{p\,2}\big), \qquad
    V_e = V_1 + V_2. \label{eq:lyapunov_1}
\end{align}

\textit{Theorem 1:} \emph{Let $g_d<0$, $g_v<0$, $r>1$ with $r^2+r g_d-g_v=0$, and $p=-\tfrac r2$. Along \eqref{est_2d:3}, for any $\omega(t)$,
\begin{equation}\label{eq:Vedot}
    \dot V_e \le -\tfrac{r}{2}\big(s_x^2+s_y^2\big) - \tfrac{|g_d g_v|}{2}\big(\tilde d_x^2+\tilde d_y^2\big) + c_a,
\end{equation}
with $c_a \triangleq a_{\max}^2\big(\tfrac{4}{r} + \tfrac{r^2}{|g_d g_v|}\big)$, and hence $\dot V_e \le -\lambda V_e + c_a$ with $\lambda \triangleq \min\{\tfrac r3,\, \tfrac{|g_d||g_v|}{|g_v|+2r^2}\}$. Consequently $\tilde e$ is globally uniformly ultimately bounded (GUUB) with ultimate bound proportional to $a_{\max}$, uniformly in $\omega$; if $a_x^{p},a_y^{p}\to0$ (e.g., constant predecessor velocity) then $\tilde e\to0$, exponentially when $a_x^{p}\equiv a_y^{p}\equiv 0$.}

\begin{proof}
Differentiating $V_1$ along the first two rows of \eqref{est_2d:3} and using $r g_d - g_v = -r^2$ gives $\dot V_1 = -r s_x^2 + g_d|g_v|\tilde d_x^2 + a_x^{p}(2s_x - r\tilde d_x) + \omega\big[-p\,s_x\tilde d_y - s_x\tilde v_y^{p} + p\,\tilde v_x^{p}\tilde d_y + \tilde v_x^{p}\tilde v_y^{p}\big]$, and symmetrically for $\dot V_2$. Summing, the $\omega$-terms collect into $(r+2p)\,\omega\,(\tilde v_x^{p}\tilde d_y - \tilde d_x\tilde v_y^{p})$, which vanishes for $p=-\tfrac r2$, leaving
\begin{align}
    \dot V_e = &-r(s_x^2+s_y^2) - |g_d g_v|(\tilde d_x^2+\tilde d_y^2) \nonumber\\
    &+ a_x^{p}(2s_x - r\tilde d_x) + a_y^{p}(2s_y - r\tilde d_y). \label{eq:Vedot_exact}
\end{align}
Young's inequality, $2a s \le \tfrac r2 s^2 + \tfrac{2a^2}{r}$ and $r a \tilde d \le \tfrac{|g_d g_v|}{2}\tilde d^2 + \tfrac{r^2 a^2}{2|g_d g_v|}$, with $|a_i^{p}|\le a_{\max}$, yields \eqref{eq:Vedot}. Since $\tilde v^{p\,2}_i \le 2r^2\tilde d_i^2 + 2s_i^2$, $V_e \le \tfrac12\sum_i\big(3s_i^2 + (|g_v|+2r^2)\tilde d_i^2\big)$, which gives $\dot V_e\le-\lambda V_e + c_a$. The comparison lemma yields $V_e(t)\le e^{-\lambda t}V_e(0) + c_a/\lambda$, i.e., GUUB by the ultimate-boundedness theorem in~\cite{khalil17}; replacing $a_{\max}$ by $\sup_{\tau\ge t}|a^{p}_i(\tau)|$ in $c_a$ shows $\tilde e\to0$ when $a^{p}_i\to0$, and $\dot V_e\le-\lambda V_e$ when $a_i^{p}\equiv0$.
\end{proof}

Denote by $\epsilon_d$ and $\epsilon_v$ the resulting ultimate bounds on $|\tilde d_i|$ and $|\tilde v_i^{p}|$; the set outside of which \eqref{eq:Vedot} is negative gives the estimates $\epsilon_d \approx \sqrt{2c_a/|g_d g_v|}$ and $\epsilon_v \approx r\epsilon_d + \sqrt{2c_a/r}$. Note that the $\omega=0$ matrix $A(0)$ has characteristic polynomial $(\lambda^2 - g_d\lambda - g_v)^2$ and is Hurwitz for $g_d,g_v<0$. A convenient parameterization satisfying $r=-2p$ and $r^2+r g_d-g_v=0$ is
\(
p=\frac{g_d}{3},\, g_v=-\frac{2g_d^2}{9},
\)
so the estimator can be tuned primarily through the single gain $g_d<0$.

\subsection{Safety-Critical Controller Design}
Leveraging the estimator described in Section~\ref{subsec:estimator_design}, each agent computes state estimates of its predecessor(s) along both $X^+$-edges and $Y$-edges as defined by the underlying DAG. In this section, we design a safety-critical tracking controller that utilizes these estimates to ensure that the follower maintains the prescribed inter-agent distances concurrently along the $X^+$- and $Y$-axes.
To enforce safety along the $X^+$-edge, we define the safety function
\begin{equation}\label{control_2d:8}
    h_1(d_x,v) \triangleq d_x - d_s - T v,
\end{equation}
where $d_x$ is the measured longitudinal distance (along the follower’s $x$-axis) from the follower to its predecessor, $d_s>0$ is the desired safe distance, $v$ is the follower's linear velocity, and $T>0$ is the prescribed time headway, which provides a buffer for reaction time and enhances string stability.

Accordingly, the safe set is defined as
\begin{equation}\label{control:safe_set1}
    \mathcal{C} \triangleq \{ (d_x,v) \in \mathbb{R}^2 : h_1(d_x,v) \geq 0 \}.
\end{equation}

\noindent\textit{Definition 1.} \emph{Let $h_1: \mathbb{R}^2 \to \mathbb{R}$ be continuously differentiable. The set $\mathcal{C}$ in (\ref{control:safe_set1}) is forward invariant if there exists an extended class $\mathcal{K}$ function~\cite{khalil17} denoted by $\alpha:\mathbb{R}\to\mathbb{R}$ such that for all $(d_x,v)$
\begin{equation}
    \dot{h}_1(d_x,v) \geq -\alpha\big(h_1(d_x,v)\big). \label{eq:CBF_condition}
\end{equation}
}

To derive an explicit bound on the follower’s acceleration that guarantees safety, we differentiate (\ref{control_2d:8}) with respect to time. Noting that
\[ \dot{d}_x = v_x^{p} + d_y\,\omega - v \quad \text{and} \quad \dot{v} = u, \]
we obtain
\begin{equation}
    \dot{h}_1 = v_x^{p} - v + d_y\,\omega - T\,u.
\end{equation}
Enforcing the CBF condition (\ref{eq:CBF_condition}) yields
\begin{equation}\label{eq:cbf_inequality}
    T\,u \leq v_x^{p} - v + d_y\,\omega + \alpha\big(h_1\big).
\end{equation}

Let $E_{vx},E_{vy}>0$ denote bounds on the velocity-estimation errors, i.e.,
$|\tilde v_x^{p}(t)|\le E_{vx}$ and $|\tilde v_y^{p}(t)|\le E_{vy}$ for all $t\ge0$.
Since $v_x^{p}$ is only available through $\hat v_x^{p}$, and $\hat v_x^{p} - E_{vx} \le v_x^{p}$, the inequality
\begin{equation}\label{eq:cbf_robust}
    T\,u \leq \hat{v}_x^{p} - E_{vx} - v + d_y\,\omega + \alpha\big(h_1\big)
\end{equation}
implies \eqref{eq:cbf_inequality}. By introducing a nonnegative tuning parameter $x_c$, we transform the inequality into the closed-form control law:
\noindent\begin{equation}\label{control:u}
    u = \kappa\Big(\hat{v}_x^{p} - E_{vx} - x_c - v + d_y\,\omega + \alpha\big(h_1\big)\Big),
\end{equation}
where $\kappa \triangleq \frac{1}{T}$.

\begin{remark}
Equation~(\ref{control:u}) is obtained by direct algebraic manipulation and therefore furnishes a \emph{closed-form} input \(u\). This avoids the quadratic program typically required, and thus removes real-time solver latency and feasibility concerns and enables deployment on low-cost embedded processors.
\end{remark}

\begin{remark}[Selecting $E_{vx},E_{vy}$]\label{rem:Ev}
The bounds $E_{vx},E_{vy}$ are not free parameters: for \eqref{eq:cbf_robust} to imply \eqref{eq:cbf_inequality} they must dominate $|\tilde v^{p}_i(t)|$ for all $t$. Theorem~1 provides this from the acceleration limit $a_{\max}$ and the gains: after a transient of duration $\lambda^{-1}\ln\!\big(\lambda V_e(0)/c_a\big)$, $|\tilde v_i^{p}|\le\epsilon_v$, while for all $t$, $|\tilde v_i^{p}|\le\sqrt{2\big(V_e(0)+c_a/\lambda\big)}$, where the initialization $\hat d_i(0)=d_i(0)$, $\hat v_i^{p}(0)=0$ gives $V_e(0)=v^{p}(0)^2\le v_{\max}^2$. Choosing $E_{vi}\ge\epsilon_v$ therefore guarantees safety after the transient, and $E_{vi}\ge\sqrt{2(V_e(0)+c_a/\lambda)}$ guarantees it for all $t$; the price of a larger $E_{vi}$ is a larger admissible setpoint in \eqref{eq:conds}, not a loss of performance. Bounded measurement noise enters \eqref{est_2d:3} as an additional bounded input and can be absorbed into $a_{\max}$. In Section~\ref{sec:results}, $E_{vi}=4\epsilon_v$.
\end{remark}

To ensure that the follower converges to the desired longitudinal position, we define the tracking error via the Lyapunov candidate
\begin{equation}\label{control_2d:6}
    V_x \triangleq \frac{1}{2}\Big(\hat{d}_x - d_x^* - T\,v\Big)^2,
\end{equation}
where $\hat{d}_x$ is the estimated longitudinal position and $d_x^*$ is the desired position. Differentiating $V_x$ along the estimator equation for $\dot{\hat d}_x$ in \eqref{est_2d:1} and substituting the control law (\ref{control:u}) yields, exactly,
\begin{align}
    \dot{V}_x &= \Big(\hat{d}_x - d_x^* - T\,v\Big)\Big(\hat{v}_x^{p} - v + d_y\,\omega + g_d\,\tilde{d}_x - T\,u\Big) \nonumber \\
    &= \Big(\hat{d}_x - d_x^* - T\,v\Big)\Big(E_{vx} + x_c - \alpha\big(h_1\big) + g_d\,\tilde{d}_x\Big).
\end{align}
Defining
\[ \hat{h}_1 \triangleq \hat{d}_x - d_s - T\,v = h_1 + \tilde d_x \quad \text{and} \quad d^*_e \triangleq d_x^* - d_s, \]
and selecting the extended class $\mathcal{K}$ function
\( \alpha(h) = -g_d\,h, \)
which is valid since $g_d < 0$, the $\tilde d_x$ terms cancel and the derivative reduces to
\begin{equation}\label{eq:Vxdot}
    \dot{V}_x = g_d\,\Big(\hat{h}_1 - d^*_e\Big)^2 + \Big(\hat{h}_1 - d^*_e\Big)\Big(E_{vx} + x_c + g_d\,d^*_e\Big).
\end{equation}

The lateral control law $\omega$ is synthesized following a symmetric procedure. Define the lateral safety function
\begin{equation}\label{control_2d:9}
    h_2 \triangleq \frac{d_s}{|d_s|}\Big(d_y-d_s\Big),
\end{equation}
where $d_y$ is the lateral distance from the relevant $Y$-edge predecessor. Using $\dot d_y = v_y^{p} - d_x\,\omega$, the CBF condition $\dot h_2 \geq -\alpha(h_2)$ with $\alpha(h)=-g_d h$, together with the conservative bound $v_y^{p}\ge \hat v_y^{p}-E_{vy}$ and a nonnegative slack variable $y_c$, yields the closed-form steering law
\begin{equation}\label{control:w}
    \omega = \frac{\hat{v}_y^{p} - g_d\big(d_y-d_s\big)}{d_x} - \frac{|d_s|\big(E_{vy}+y_c\big)}{d_s\,d_x}.
\end{equation}
The division by $d_x$ is well-posed because the longitudinal CBF enforces $d_x \ge d_s + Tv \ge d_s > 0$.

To regulate the lateral offset, define
\begin{equation}\label{control_2d:10}
    V_y \triangleq \frac{1}{2}\Big(\hat d_y-d_y^*\Big)^2.
\end{equation}
Substituting \eqref{control:w} into $\dot V_y = (\hat d_y-d_y^*)\dot{\hat d}_y$ gives, again exactly,
\begin{equation}\label{eq:Vydot}
    \dot V_y {=}
    g_d\,\big(\hat d_y{-}d_y^*\big)^2
    + \big(\hat d_y{-}d_y^*\big)
    \Big(
        \frac{|d_s|\big(E_{vy}{+}y_c\big)}{d_s}
        + g_d\big(d_y^*{-}d_s\big)
    \Big).
\end{equation}

\begin{remark}[Input constraints]\label{rem:sat}
The laws \eqref{control:u} and \eqref{control:w} are applied with saturation to the limits in \eqref{bnd:1}; $\omega$ is computed and saturated first, and its applied value is used in \eqref{control:u}. Each CBF constraint is a \emph{one-sided} bound on its input: an upper bound on $u$ in \eqref{eq:cbf_robust}, and an upper (lower) bound on $\omega$ for $d_s>0$ ($d_s<0$). Saturating on the constrained side, e.g., $u=u_{\max}$, therefore cannot violate it, and saturating on the opposite side preserves it whenever the required deceleration or turn rate is within the actuator authority, i.e., $-Tu_{\max}\le \hat{v}_x^{p} - E_{vx} - v + d_y\,\omega + \alpha(h_1)$ for the longitudinal channel. We assume this feasibility condition holds, as it did in all experiments; certifying it a priori through a braking-aware $h_1$ is left for future work.
\end{remark}

\begin{remark}[Pairwise coverage]\label{rem:pairs}
Forward invariance is established per edge of $G$. The edge sets in Figs.~\ref{fig:setup_2d} and~\ref{fig_gazebo_snapshot} are chosen so that every robot has a constraint with each robot it can approach under the desired geometry; pairs not joined by an edge (e.g., $F_3$--$L$ in Fig.~\ref{fig:setup_2d}) are separated by the offsets of the intermediate edges and are not covered by an explicit guarantee. Full pairwise coverage can be obtained by adding edges, each contributing one further one-sided bound of the form \eqref{eq:cbf_robust}, with the most restrictive bound applied.
\end{remark}

We now combine \eqref{eq:lyapunov_1}, \eqref{control_2d:6}, and \eqref{control_2d:10} into the composite Lyapunov function
\begin{equation}\label{lyapunov_2}
    V \triangleq V_x + V_y + V_e ,
\end{equation}
and write $e_x \triangleq \hat h_1 - d_e^*$ and $e_y \triangleq \hat d_y - d_y^*$ for the tracking errors in the estimated coordinates. The slack parameters are fixed as
\begin{equation}\label{eq:xc_yc}
    x_c = -g_d\,(d_x^*-d_s) - E_{vx}, \quad y_c = \frac{|d_s|}{d_s}\big(-g_d\,(d_y^*-d_s)\big) - E_{vy},
\end{equation}
which are nonnegative exactly when
\begin{equation}\label{eq:conds}
    d_x^* \geq d_s - \frac{E_{vx}}{g_d} \quad \text{and} \quad
    \begin{cases}
        d_y^* \geq d_s - \frac{E_{vy}}{g_d} & \text{if } d_s > 0,\\[0pt]
        d_y^* \leq d_s + \frac{E_{vy}}{g_d} & \text{if } d_s < 0,
    \end{cases}
\end{equation}
i.e., when the setpoints are compatible with the safety margins.

\textit{Theorem 2:}  \emph{Consider the estimator \eqref{est_2d:1}, the control laws \eqref{control:u} and \eqref{control:w} with \eqref{eq:xc_yc}, and assume \eqref{eq:conds} and $|\tilde v_i^{p}(t)|\le E_{vi}$ for all $t$ (Remark~\ref{rem:Ev}). Then (a) the safe sets $\{h_1\ge0\}$ and $\{h_2\ge0\}$ are forward invariant for any predecessor motion; and (b) if the predecessor maintains constant velocity ($a_x^{p}=a_y^{p}=0$), then $d_x \to d_x^* + T\,v$ and $d_y \to d_y^*$ exponentially.}

\begin{proof}
(a) Since $x_c\ge0$ by \eqref{eq:conds} and $|\tilde v_x^{p}|\le E_{vx}$, \eqref{control:u} satisfies \eqref{eq:cbf_robust} and hence \eqref{eq:cbf_inequality}, which is the CBF condition \eqref{eq:CBF_condition}; forward invariance of $\{h_1\ge0\}$ follows from~\cite{ames2019control}. The same argument with $y_c\ge0$ applies to $h_2$.
(b) Substituting \eqref{eq:xc_yc} into \eqref{eq:Vxdot} and \eqref{eq:Vydot} cancels the linear terms, so with \eqref{eq:Vedot} at $a_{\max}=0$,
\begin{equation}\label{eq:Vdot_final}
    \dot{V} \le g_d\,e_x^2 + g_d\,e_y^2 - \lambda V_e ,
\end{equation}
which is negative definite in $(e_x,e_y,\tilde e)$ because $g_d<0$ and $\lambda{>}0$. Hence $e_x,e_y,\tilde e\to0$ exponentially, and since $d_x {-} d_x^* - Tv = e_x {-} \tilde d_x$ and $d_y {-} d_y^* = e_y - \tilde d_y$, the claim follows.
\end{proof}

\textit{Theorem 3:}  \emph{Under the assumptions of Theorem~2 but with time-varying predecessor motion, $|a^{p}_i|\le a_{\max}$, the estimated tracking errors satisfy $\dot e_x = g_d e_x$ and $\dot e_y = g_d e_y$ and converge to zero exponentially, all states are GUUB, and the true formation errors are ultimately bounded by the estimator bound:
\begin{equation}\label{eq:tv_bounds}
    \limsup_{t\to\infty}|d_x - d_x^* - T\,v| \le \epsilon_d, \qquad
    \limsup_{t\to\infty}|d_y - d_y^*| \le \epsilon_d .
\end{equation}}

\begin{proof}
With \eqref{eq:xc_yc}, \eqref{eq:Vxdot} and \eqref{eq:Vydot} read $e_x\dot e_x = g_d e_x^2$ and $e_y\dot e_y = g_d e_y^2$, i.e., $\dot e_i = g_d e_i$; notably, the estimated tracking dynamics are decoupled from $\tilde e$ and from $a^{p}_i$, so $e_i\to0$ exponentially regardless of predecessor motion. By Theorem~1, $\tilde e$ is GUUB with $\limsup|\tilde d_i|\le\epsilon_d$, so $\dot V\le g_d(e_x^2+e_y^2)-\lambda V_e + c_a$ is negative outside a compact set and all states are GUUB~\cite{khalil17}. The bounds \eqref{eq:tv_bounds} follow from $d_x - d_x^* - Tv = e_x - \tilde d_x$ and $d_y - d_y^* = e_y - \tilde d_y$.
\end{proof}

Theorem~3 makes precise the distinction between exact convergence and ultimate boundedness: what converges exactly is the tracking error in the estimated coordinates, while the true offsets converge to a neighborhood whose size $\epsilon_d$ is set by the predecessor's acceleration limit and shrinks with the estimator gains; safety, in contrast, holds exactly by Theorem~2(a). Although the analysis is based on a kinematic model that neglects friction and inertia, the approach can be extended to more complex dynamics by incorporating the corresponding physical effects.

\subsection{String Stability}\label{subsec:string}

The safety function \eqref{control_2d:8} and the setpoint $d_x\to d_x^*+Tv$ implement the constant time-headway spacing policy $d_x\ge d_s+Tv$, which Swaroop and Hedrick~\cite{swaioa94} identified as the structural requirement for string stability of platoons and which we also employed in~\cite{TVT2024}. Time headway alone does not guarantee attenuation, however, since the closed loop here also contains the estimator. The following result gives an explicit condition.

\textit{Proposition 1:} \emph{Consider a follower and its $X^+$-predecessor moving along a straight line ($\omega=\omega^{p}=0$) under \eqref{control:u} with \eqref{eq:xc_yc}. Up to the exponentially decaying term $e_x$, the follower's velocity responds to the predecessor's velocity through
\begin{equation}\label{eq:G}
    G(s) = \frac{|g_d|\,s + |g_v|}{\big(s^2 + |g_d|\,s + |g_v|\big)\big(Ts+1\big)},
\end{equation}
and $\mathcal{S}=\|G\|_{\mathcal{H}_\infty}\le1$ whenever $T^2|g_v|\ge2$ and $g_d^2\ge2|g_v|$. Under the parameterization $g_v=-2g_d^2/9$ these reduce to $T|g_d|\ge3$. Along any predecessor chain of $G$, the gain from the leader to the $k$-th follower is then at most $1$.}

\begin{proof}
From $e_x = \hat d_x - d_x^* - Tv$ and $\hat d_x = d_x + \tilde d_x$, $T\dot v = (v^{p} - v) + \dot{\tilde d}_x - g_d e_x$. With $\omega=0$ the $x$-channel of \eqref{est_2d:3} gives $\tilde D_x(s) = -s\,V^{p}(s)/(s^2 - g_d s - g_v)$, and substituting yields \eqref{eq:G}. A direct computation gives $|G(j\varpi)|^2\le1$ iff $\varpi^2\big[\varpi^2 - 2|g_v| + T^2\big((|g_v|-\varpi^2)^2 + g_d^2\varpi^2\big)\big]\ge0$; the bracket is nondecreasing in $\varpi^2$ when $g_d^2\ge2|g_v|$ and equals $|g_v|(T^2|g_v|-2)\ge0$ at $\varpi=0$. The chain statement follows because $\mathcal{H}_\infty$ gains multiply along a cascade.
\end{proof}

Proposition~1 covers the longitudinal channel; curved motion and the lateral coupling through $d_y\omega$ in \eqref{control:u} are not covered analytically and are evaluated empirically in Section~\ref{sec:results}. With the gains used there ($T=0.2$~s, $g_d=-15$) the condition $T|g_d|\ge3$ holds with equality.

\section{Results and Evaluation}\label{sec:results}

This section evaluates the proposed estimator and control law across several formation control scenarios: estimator accuracy and convergence with zero and non-zero angular velocity; a four-agent diamond formation (Fig.~\ref{fig:setup_2d}) with a constant-velocity leader; disturbance propagation through the formation; and, beyond Python simulation, a three-agent triangular formation in the physics engine-based Gazebo environment, which introduces realistic physics and sensor noise.\looseness=-1

The continuous-time behavior of the system is simulated using a 4th-order Runge-Kutta method, while the distance sensor data is generated by computing the state differences between agents. The following bounds are imposed on the agents:\looseness=-1
\[ v \leq 1\;\text{m/s}, \quad u \leq 0.5\;\text{m/s}^2, \quad \omega \leq 2\;\text{rad/s}. \]
The estimator gain is set to
\( g_d = -15 \),
with the remaining parameters computed as
\[ g_v = -50, \quad p = -5, \quad r = 10, \quad T = 0.2\,\text{sec}, \quad \lambda = 3. \]
With these gains, $A(0)$ in~(\ref{est_2d:3}) has eigenvalues $\{{-}5,\!{-}5,\!{-}10,\!{-}10\}$. For the acceleration levels exercised by the leader profiles ($a_{\max}=0.5$~m/s$^2$), Theorem~1 gives $c_a=0.13$, $\epsilon_d\approx0.019$~m, and $\epsilon_v\approx0.35$~m/s; following Remark~\ref{rem:Ev}, the safety bounds are set to $E_{vx}\!=\!E_{vy}\!=\!4\epsilon_v\approx1.4~\text{m/s}$ (all agents start from rest, so $V_e(0)=0$ and the transient bound is inactive).

\subsection{Estimator Analysis}

We first benchmark the estimator in a two-robot simulation. Both agents accelerate for $0$--$2$ seconds at $0.2$ m/s$^2$; Agent 1 rotates at $\omega\!=\!0.2$ rad/s while Agent 2 keeps $\omega=0$.
Three cases are analyzed:
\begin{itemize}\setlength{\itemsep}{0pt}\itemindent=-5pt
  \item[a)] \textit{Acceleration phase} ($0$--$2$ sec): Agent 1 estimates the still-accelerating Agent 2.
  \item[b)] \textit{Cruise, rotating observer} ($t>2$ sec): Agent 1 (rotating) estimates Agent 2 cruising at constant speed.
  \item[c)] \textit{Cruise, rotating target} ($t>2$ sec): Agent 2 (no rotation) estimates the cruising, rotating Agent 1.
\end{itemize}

\begin{figure}[tbp]
    \centering
    \renewcommand{\arraystretch}{0.5}
    \setlength{\tabcolsep}{0pt}
    \begin{tabular}{cc}
    \includegraphics[width=0.4\columnwidth]{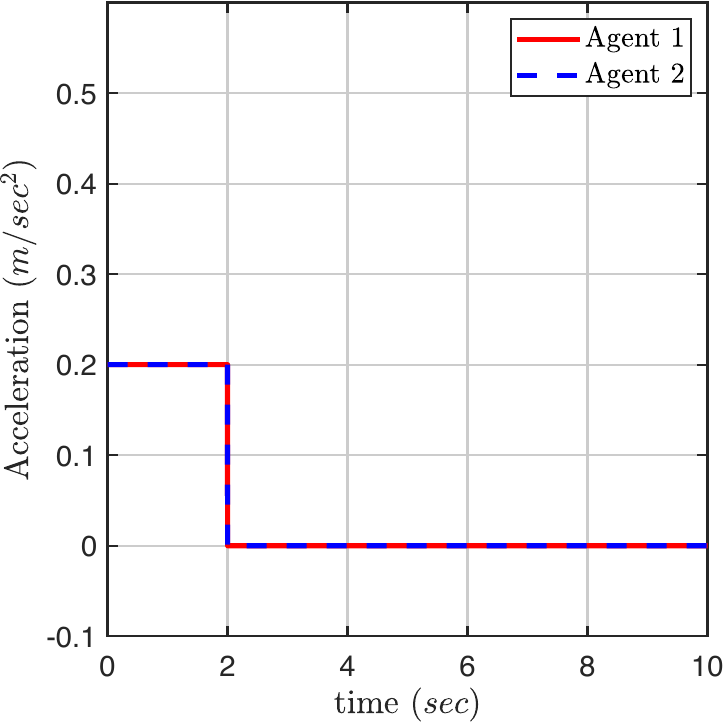} &
    \includegraphics[width=0.4\columnwidth]{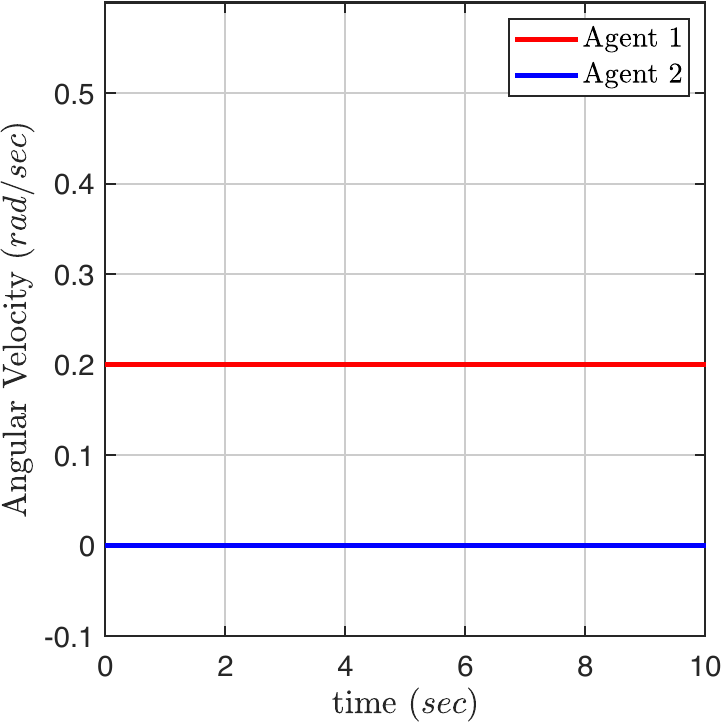}
    \tabularnewline
    \small{(a)} & \small{(b)}
    \tabularnewline
    \includegraphics[width=0.4\columnwidth]{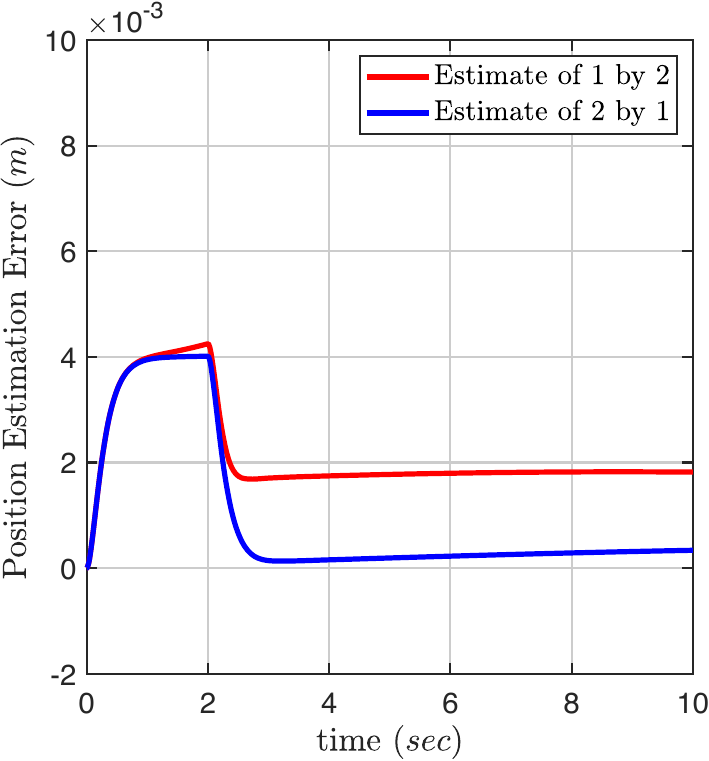} &
    \includegraphics[width=0.4\columnwidth]{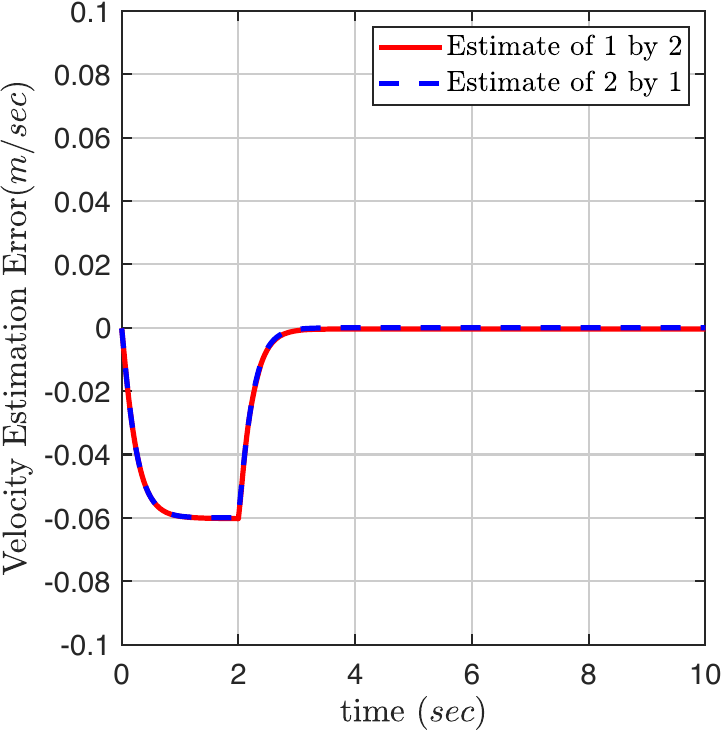}
    \tabularnewline
    \small{(c)} & \small{(d)} \\[4pt]
    \end{tabular}
    \caption{Estimator performance for a two-agent system with different control profiles: (a) Linear acceleration profiles; (b) Angular velocities; (c) Position estimation errors; (d) Velocity estimation errors.}
    \label{fig:2d_1}
\end{figure}

For scenario (a), during the first 2 seconds both agents accelerate at 0.2 m/s$^2$.
Consistent with Theorem~1, the estimator is expected to exhibit GUUB behavior, with position and velocity errors within $\epsilon_d\approx0.019$~m and $\epsilon_v\approx0.35$~m/s. Numerical simulation results in Fig.~\ref{fig:2d_1} show that the position error converges to about 0.004 m and the velocity error to about 0.06 m/s, both well within the expected bounds.

In scenario (b), after the acceleration phase, Agent 2 moves at a constant velocity while Agent 1 maintains its angular motion. Since the estimator's convergence is independent of the observer's angular velocity (Theorem~1), the estimation errors are anticipated to approach zero. Simulation results in Fig.~\ref{fig:2d_1} confirm this behavior, with the distance error converging to approximately 0.0003 m and the velocity error to roughly 0.00004 m/s.

In scenario (c), Agent 1 continues with constant linear and angular velocity, while Agent 2 remains with zero angular motion. Here, the estimator effectively interprets the angular motion as an additional acceleration term, computed as $a = u + v\,\omega = 0.08$ m/s$^2$. Accordingly, the estimation errors are expected to remain GUUB with bounds similar to those in scenario (a). The simulation results in Fig.~\ref{fig:2d_1} corroborate this, showing convergence of the distance error to approximately 0.0018 m and the velocity error to about 0.0004 m/s.

\subsection{Formation Control with Constant Leader Velocity}\label{sec:results_2d_2}

\begin{figure}[tbp]
    \centering
    \renewcommand{\arraystretch}{0.4}
    \setlength{\tabcolsep}{0pt}
    \begin{tabular}{cc}
    \includegraphics[width=0.44\columnwidth]{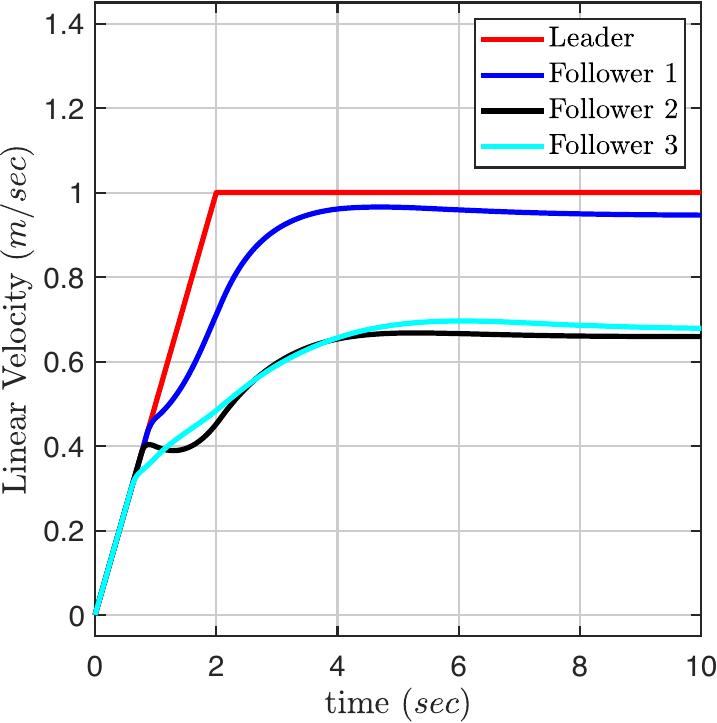} &
    \includegraphics[width=0.44\columnwidth]{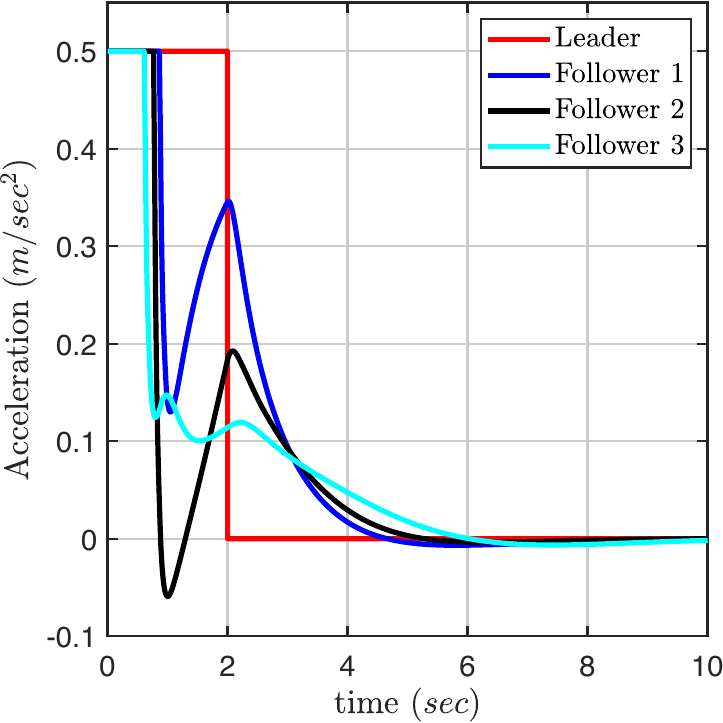} \\
    \small{(a)} & \small{(b)} \\
    \includegraphics[width=0.44\columnwidth]{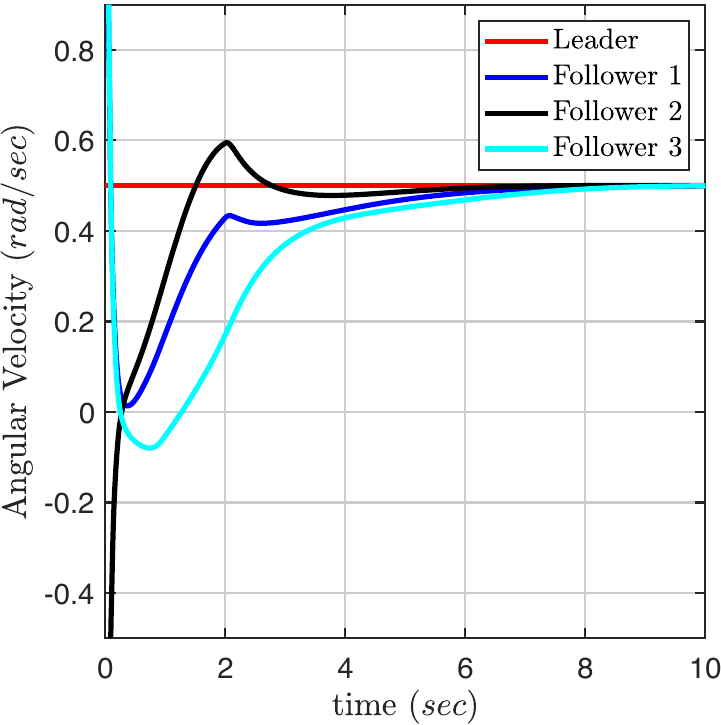} &
    \includegraphics[width=0.44\columnwidth]{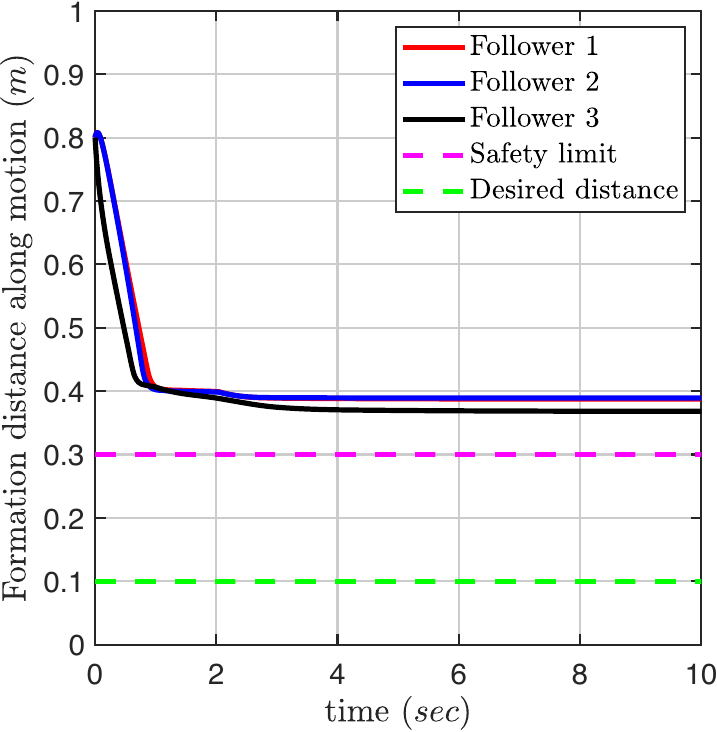} \\
    \small{(c)} & \small{(d)} \\
    \includegraphics[width=0.44\columnwidth]{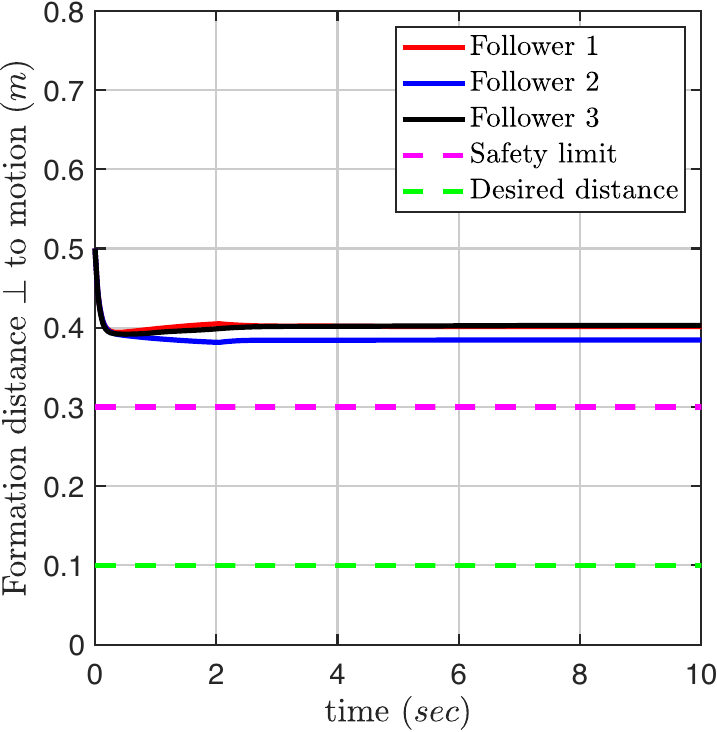} &
    \includegraphics[width=0.43\columnwidth]{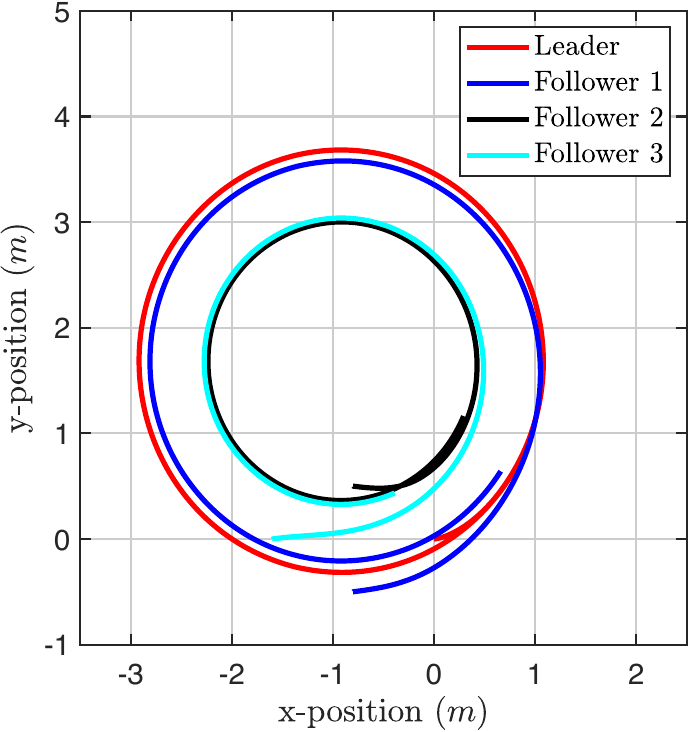} \\
    \small{(e)} & \small{(f)} \\[4pt]
    \end{tabular}
    \caption{Formation control for circular motion using the proposed framework: (a) Linear velocities; (b) Accelerations; (c) Angular velocities; (d) Distance along the motion; (e) Distance perpendicular to the motion; (f) Trajectories.}
    \label{fig:2d_2}
\end{figure}

This section evaluates the proposed control framework under a constant leader velocity scenario. The leader agent accelerates to a fixed linear velocity while maintaining a constant angular velocity of $\omega = 0.5$ rad/s, thereby steering along a circular trajectory, which is a common benchmark in formation control studies. Due to the non-zero angular velocity, both acceleration components $a_x^{p}$ and $a_y^{p}$ are non-zero, situating this scenario within the time-varying analysis of \textit{Theorem~3}.\looseness=-1

Figures~\ref{fig:2d_2}(a), (b), and (c) display the linear velocity, acceleration, and angular velocity profiles for each agent throughout the experiment. The results indicate rapid convergence to a circular formation that closely tracks the leader’s trajectory. In particular, the angular velocities of the follower agents converge to that of the leader, and their linear velocities adjust according to their respective circular motion radii. This coordinated behavior is achieved entirely through the proposed estimator-controller framework without any modifications to the control law.

The safety region for each agent is defined as 0.3 m, with a desired separation of 0.1 m between followers. Figures~\ref{fig:2d_2}(d) and (e) illustrate the evolution of the safety control barrier functions along and perpendicular to the agents' motion, respectively. Despite the desired separation falling within what might conventionally be considered an unsafe region, the agents consistently maintain a conservative minimum distance as prescribed by the safety constraints. Finally, Fig.~\ref{fig:2d_2}(f) visually depicts the overall formation motion during the experiment.

A supplementary video demonstrating the circular formation scenario in Fig.~\ref{fig:2d_2} is provided alongside this paper.

\subsection{String Stability Performance}

\begin{figure}[tbp]
    \centering
    \setlength{\tabcolsep}{0pt}
    \begin{tabular}{cc}
    \includegraphics[width=0.45\columnwidth]{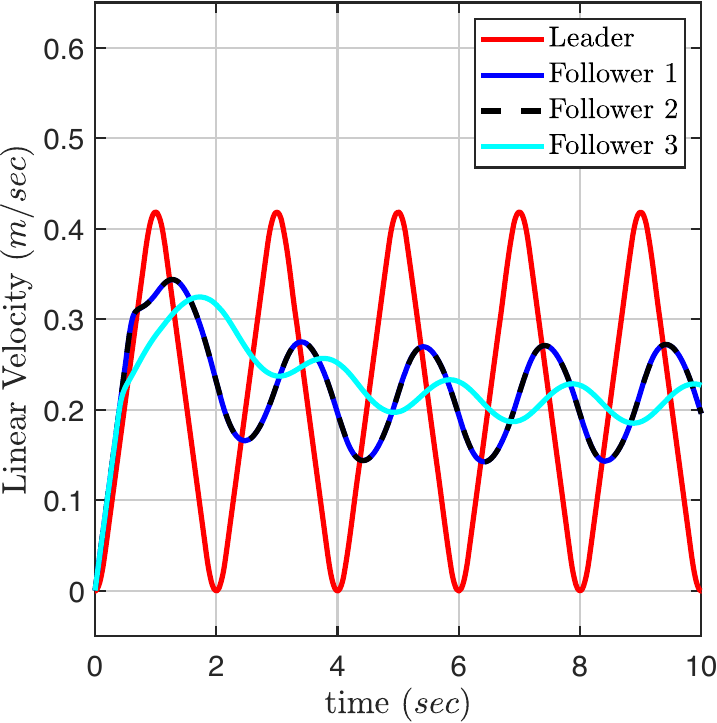} &
    \includegraphics[width=0.45\columnwidth]{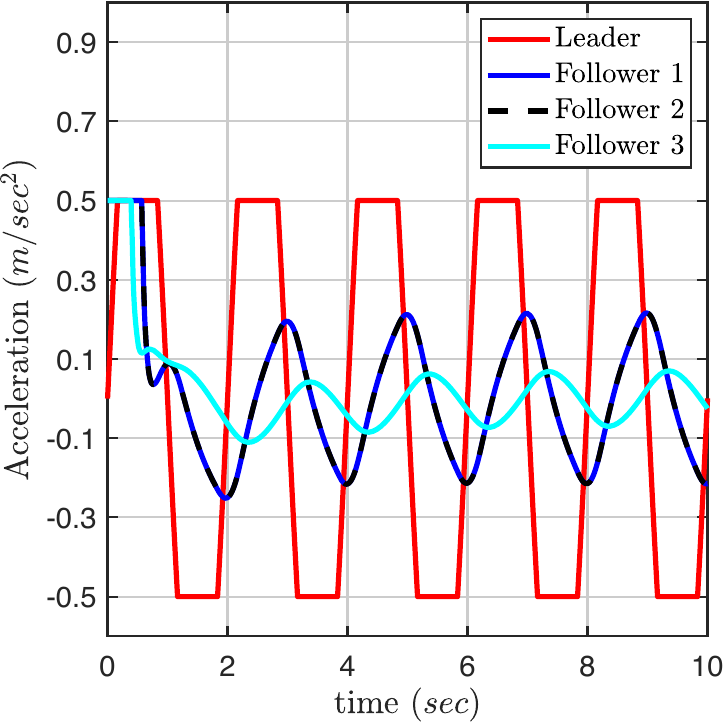}
    \end{tabular}
    \caption{Disturbance attenuation in linear velocity and acceleration for a 4-agent diamond formation control. Left: linear velocity disturbance of 0.415 m/s. Right: acceleration disturbance of 1.0 m/s$^2$.}
    \label{fig:2d_1_1}
\end{figure}

String stability ensures that disturbances, such as sudden braking or acceleration, do not amplify as they propagate from the leader to the followers, thereby reducing the risk of collisions. In a string-stable formation, the amplitude of sinusoidal velocity disturbances should decrease along the formation. Fig.~\ref{fig:2d_1_1} illustrates the performance of the proposed controller in this respect for the diamond formation of Fig.~\ref{fig:setup_2d}, which exercises curved motion, lateral coupling, and multiple branches beyond the scope of Proposition~1.

In this experiment, the leader's velocity underwent a disturbance with an amplitude of 0.415 m/s. As the disturbance propagated through the formation, the resulting amplitudes diminished to 0.12 m/s for both follower 1 and follower 2, and further to 0.035 m/s for follower 3, as shown in Fig.~\ref{fig:2d_1_1}(a). Similarly, when an acceleration disturbance of 1.0 m/s$^2$ was introduced, the proposed framework yielded acceleration responses of 0.434 m/s$^2$ for both follower 1 and follower 2, and 0.14 m/s$^2$ for follower 3, as shown in Fig.~\ref{fig:2d_1_1}(b). Notably, the overlapping profiles of follower 1 and follower 2 are attributed to their identical state evolution and pursuit of the same predecessor.

The measured per-edge amplitude ratios are $\mathcal{S}\le0.29$ for velocity and $\mathcal{S}\le0.44$ for acceleration, well below the bound $\mathcal{S}\le1$ of Proposition~1, confirming attenuation along the formation.
A clip of the oscillatory velocity disturbance experiment in Fig.~\ref{fig:2d_1_1} is provided in the video submitted with the paper.

\subsection{Physics Engine-Based Experiments}

\begin{figure}[t]
    \centering
    \begin{tikzpicture}
        \node (img) {\includegraphics[trim={0 0 0cm 0},clip,width=0.48\textwidth]{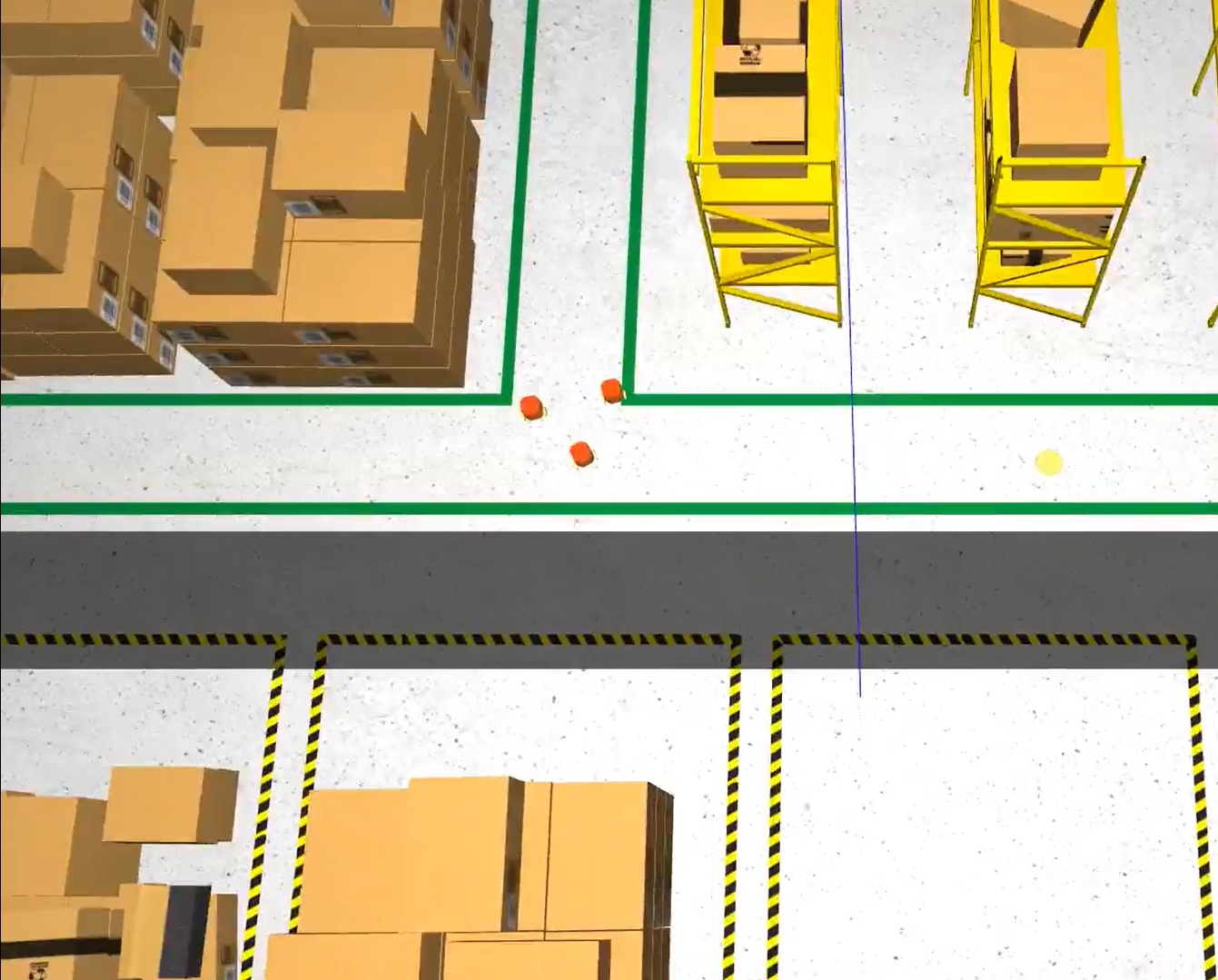}};
        \node [align=center] at (-1.9,0.45){\footnotesize Follower 1};
        \node [align=center] at (1.425,0.85){\footnotesize Follower 2};
        \node [align=center] at (0.925,0.275){\footnotesize Leader};
        \draw[dashed] (-1.2,0.5)--(-0.7,0.5);
        \draw[dashed] (0.7,0.8)--(0.25,0.8);
        \draw[dashed] (0.45,0.25)--(-0.05,0.25);
    \end{tikzpicture}
    \vspace{-15pt}
    \caption{Snapshot of the warehouse environment in Gazebo with three Turtlebot3 in a triangular formation.}
    \label{fig_gazebo_snapshot}
\end{figure}
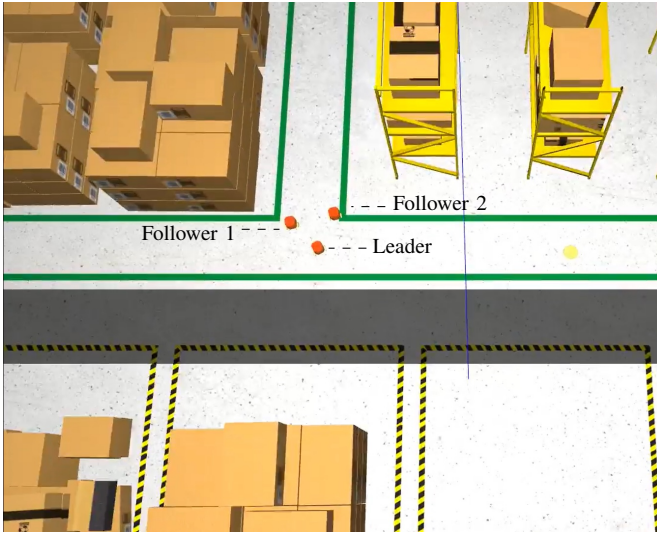

To assess the feasibility and effectiveness of the proposed estimation-based controller in realistic scenarios, we implement the framework within the Gazebo simulation environment. In this experiment, a leader agent follows a time-varying trajectory within a simulated warehouse while two follower agents maintain a triangular formation with respect to the leader. Turtlebot3 models are employed to emulate the robotic agents. Key parameters for robot motion are set as follows: maximum linear velocity of 0.3 m/s, maximum angular velocity of 1.0 rad/s, and maximum acceleration of 1.0 m/s$^2$. Figure~\ref{fig_gazebo_snapshot} shows a snapshot of the simulated warehouse environment.

A three-agent cluster operates without inter-vehicle communication. In this configuration, the leader exhibits variable velocity, and the follower agents maintain formation using the proposed controller. Initially, follower 1 and follower 2 are positioned at coordinates $(-0.5, -0.4)$ and $(-0.5, 0.4)$, respectively, relative to the leader. Emulated distance sensors provide measurements based on state differences between the agents. The estimator and controller are configured with parameters $g_d = -6$, $g_v = -8$, $p = -2$ (i.e., $r=4$), $E_{vx} = E_{vy} = 0.4~\text{m/s}$ (Remark~\ref{rem:Ev}, with $a_{\max}\approx0.1$~m/s$^2$ exercised by the leader), and $T = 0.2$ sec. Under these settings, $A(0)$ has eigenvalues $[-2,\,-2,\,-4,\,-4]$. Safety constraints are set to maintain a minimum distance of 0.2 m in all directions. In the formation, follower 1 and follower 2 track the leader along both the $X^+$ and $Y$ edges. Their formation parameters are configured such that follower 1 maintains 0.4 m longitudinally and 0.3 m laterally, while follower 2 maintains 0.4 m longitudinally and -0.3 m laterally from the leader.\looseness=-1

\begin{figure}[tbp]
    \centering
    \renewcommand{\arraystretch}{0.01}
    \setlength{\tabcolsep}{0pt}
    \begin{tabular}{cc}
    \includegraphics[width=0.46\columnwidth]{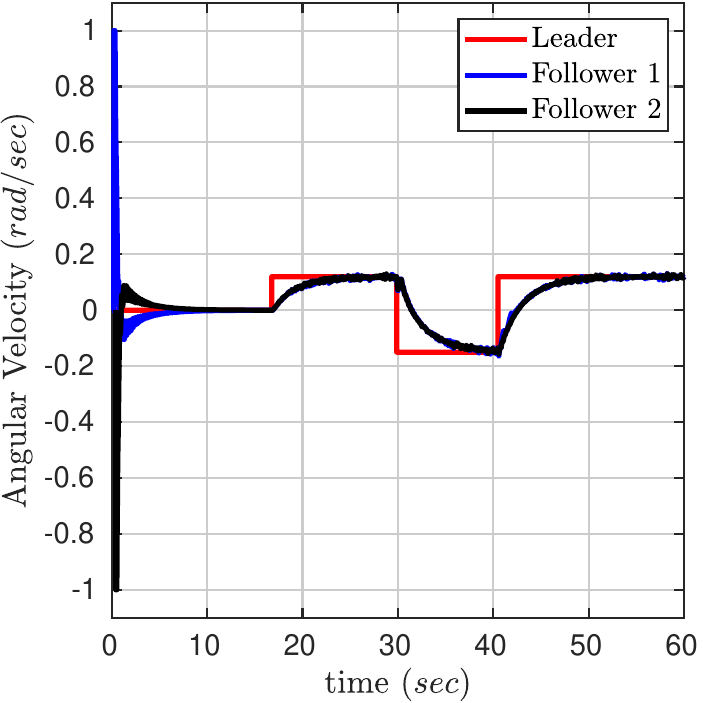} &
    \includegraphics[width=0.46\columnwidth]{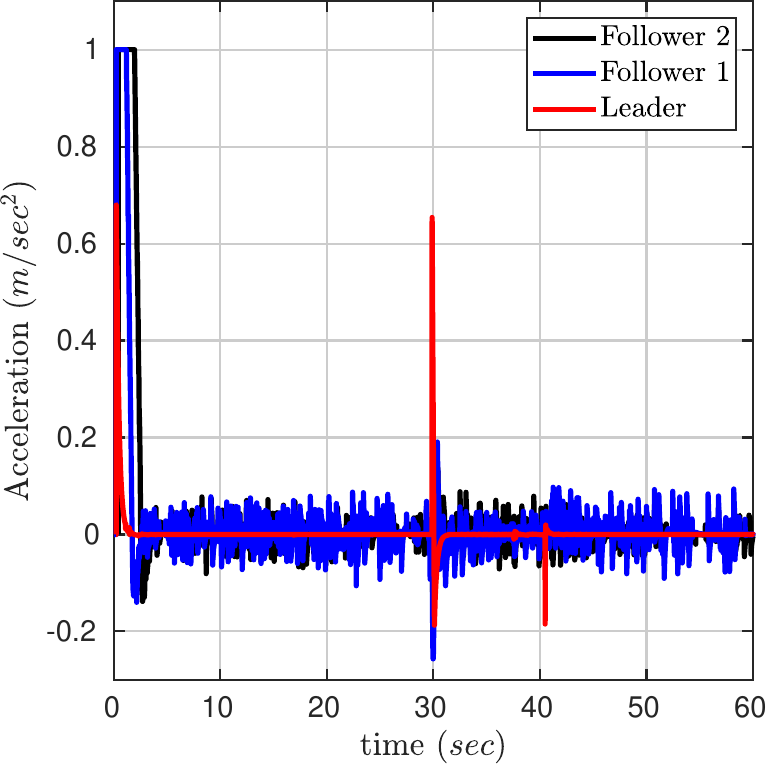} \\
    \footnotesize{(a)} & \footnotesize{(b)} \\
     \\
    \includegraphics[width=0.46\columnwidth]{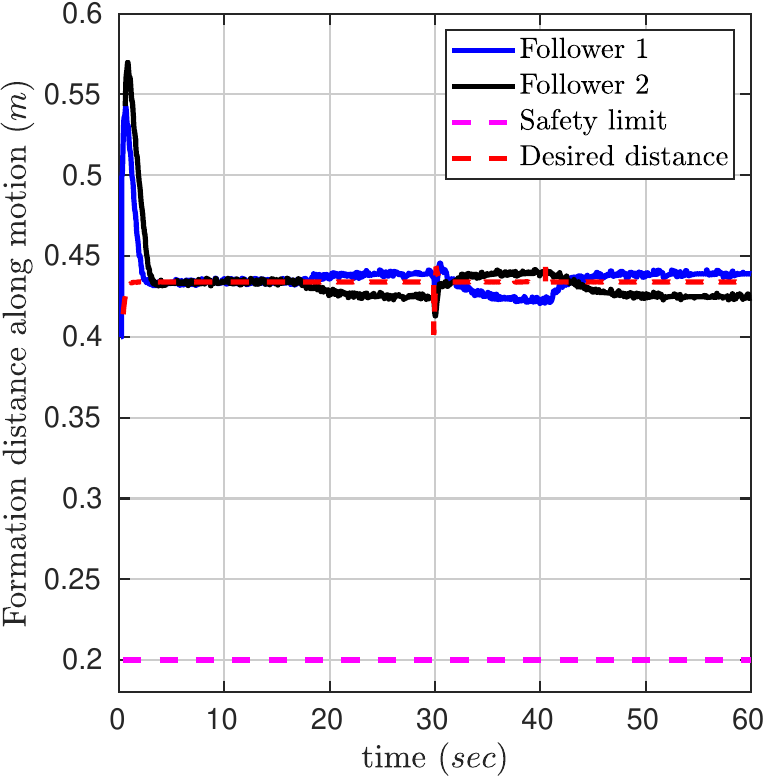} & \includegraphics[width=0.46\columnwidth]{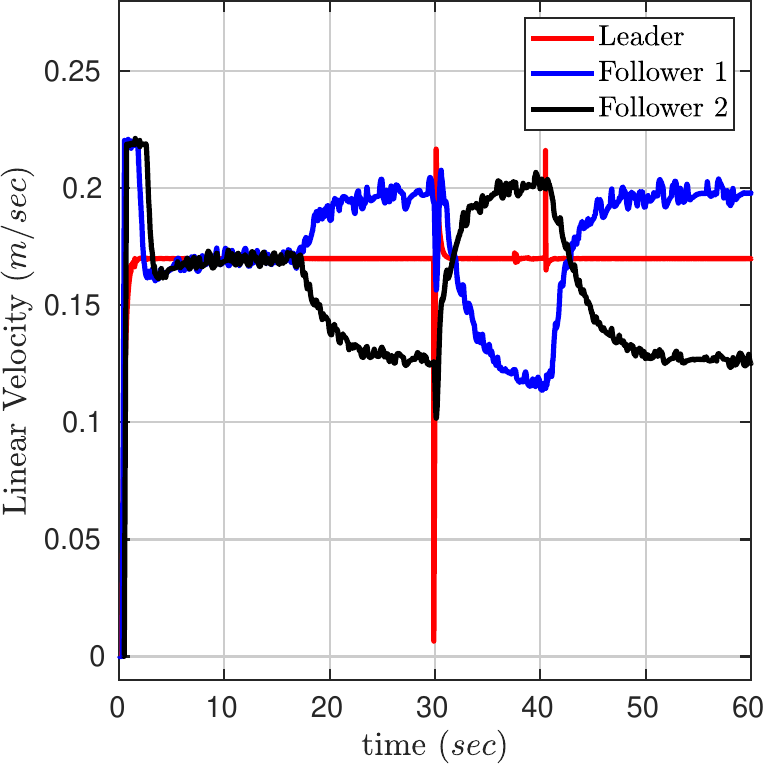} \\
    \footnotesize{(c)} & \footnotesize{(d)} \\
     \\
    \includegraphics[width=0.46\columnwidth]{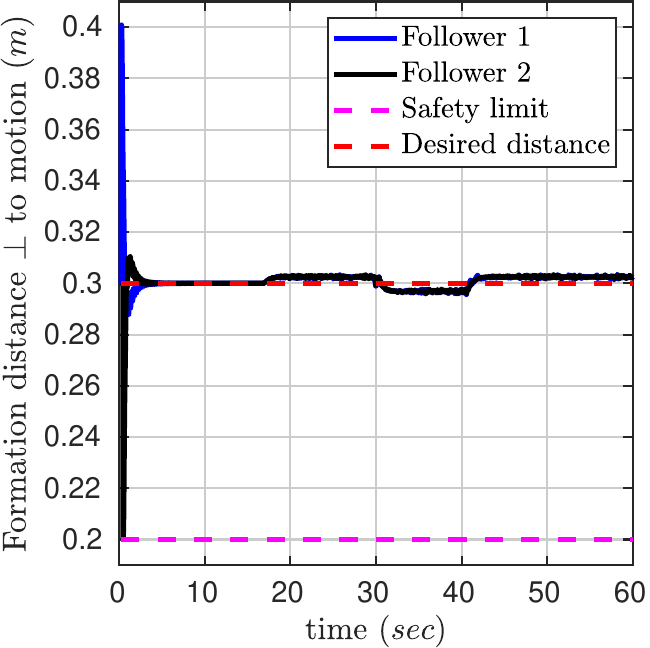} &
    \includegraphics[width=0.46\columnwidth]{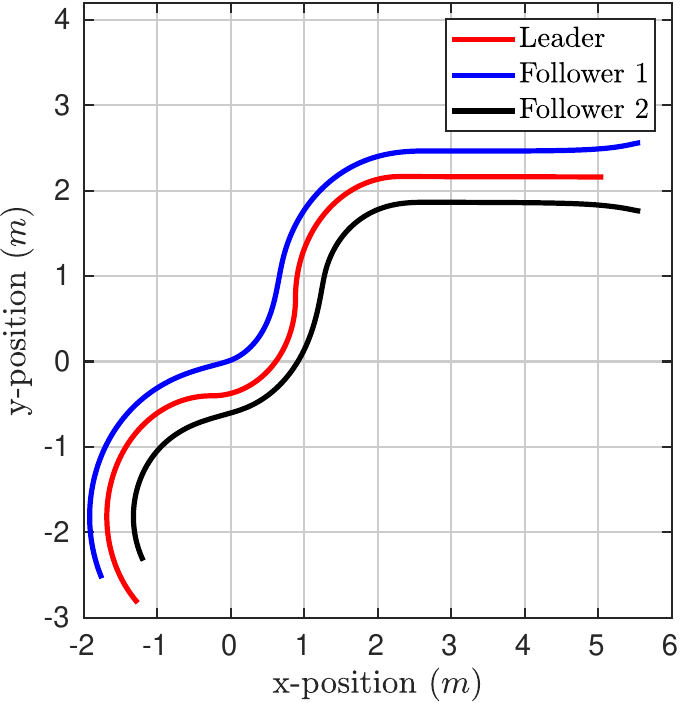} \\
    \footnotesize{(e)} & \footnotesize{(f)} \\[4pt]
    \end{tabular}
    \caption{Plots for formation control of triangular formation in Gazebo environment using the proposed framework: (a) Angular Velocity, (b) Acceleration, (c) Linear Velocity, (d) Distance along the motion, (e) Distance perpendicular to the motion, and (f) Overall motion of the formation.}
    \label{fig:gazebo_state_plots}
\end{figure}

The leader agent's time-varying trajectory is determined by the angular velocity and acceleration curves shown in Fig.~\ref{fig:gazebo_state_plots}(a) and (b). These plots indicate that the proposed controller effectively generates the necessary acceleration and angular velocity commands to maintain the desired triangular formation. This is further supported by Fig.~\ref{fig:gazebo_state_plots}(c), which illustrates the evolution of the agents' linear velocities. When the leader operates with zero angular velocity, the linear velocities of both follower agents converge to that of the leader. Conversely, when the leader rotates about a center of rotation, the follower closer to the center reduces its speed, while the follower farther away increases its speed to preserve formation integrity. Notably, the angular velocities of all agents remain synchronized, as demonstrated in Fig.~\ref{fig:gazebo_state_plots}(a).\looseness=-1

Figures~\ref{fig:gazebo_state_plots}(d) and (e) illustrate that the formation distances converge to their desired values when the leader's acceleration and angular velocity are zero (Theorem~2). In scenarios with non-zero angular velocity, deviations between the actual formation distances and the desired distances remain bounded within 0.02 m along the direction of motion and 0.005 m in the perpendicular direction, as predicted by Theorem~3. Importantly, both follower agents consistently adhere to the predefined safety limits throughout the experiment. Finally, Fig.~\ref{fig:gazebo_state_plots}(f) visually represents the trajectories of the leader and follower agents.

\begin{figure}[tbp]
    \centering
    \vspace{-8pt}
    \setlength{\tabcolsep}{0pt}
    \begin{tabular}{cc}
    \includegraphics[width=0.45\columnwidth]{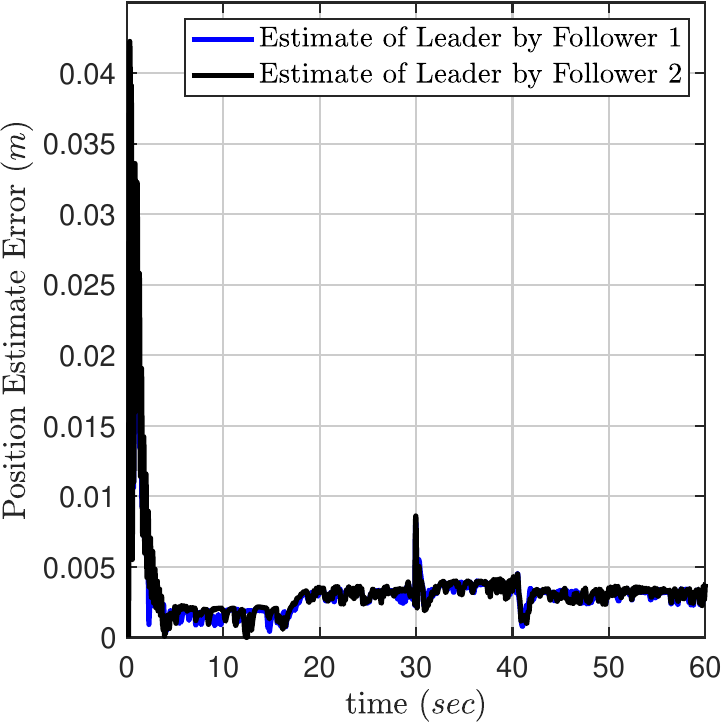} &
    \includegraphics[width=0.45\columnwidth]{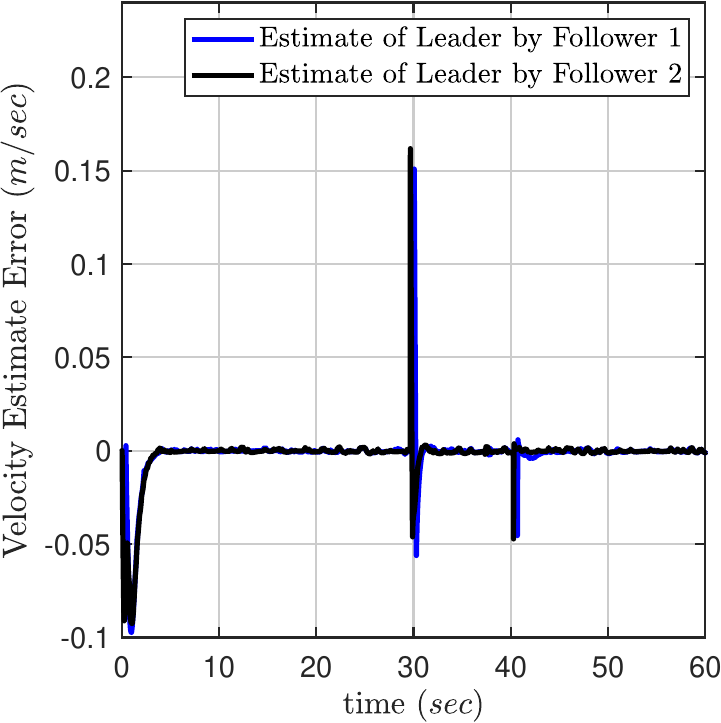}
    \end{tabular}
    \caption{Leader's position (left) and velocity (right) estimation error when estimated by follower agents.}
    \label{fig:gazebo_estimate_error}
\end{figure}

The stability of the estimator within the Gazebo environment is critical for achieving optimal performance of the proposed framework. Fig.~\ref{fig:gazebo_estimate_error} illustrates the estimation errors in both the position and velocity of the leader as measured by the follower agents. As the followers track the leader, the estimation errors are maintained at low levels; specifically, when the leader operates with zero angular velocity, the position estimation error remains below 0.002 m. For scenarios with non-zero leader angular velocity, the position estimation errors for both followers are bounded within 0.004 m. Moreover, the velocity estimation errors for both follower agents are kept below 0.006 m/s, which underscores the estimator's robustness in the presence of noise in real-world environments.

Videos of the experiments presented in this section, including a video of the Gazebo-based triangular formation experiments, can be found in the accompanying video for the article, available at: \;\texttt{\small \url{vimeo.com/1075016147/41612f2f8c}}

\section{Conclusions}\label{sec:conclusion}

This paper addressed safety-critical formation control for non-holonomic multi-robot systems under communication constraints. The proposed framework integrates sensing, estimation, and control through a decentralized estimator-based controller that achieves deterministic, real-time performance without explicit inter-agent communication.

The approach delivers several key advantages for real-world applications:  (i) closed-form control laws with microsecond level computation, avoiding the latency and feasibility issues of optimization-based methods; (ii) communication free operation that remains effective under RF constraints; (iii) safety guarantees via CBFs whose robustness margins are derived from the estimator analysis rather than tuned empirically; and (iv) an explicit string stability condition that prevents disturbance amplification across the formation. Validation in simulation and physics-based experiments indicates readiness for deployment on real robotic platforms. Future work will certify input feasibility through braking-aware barrier functions, extend safety coverage to all robot pairs in larger formations, and validate the framework on physical hardware.

\bibliographystyle{IEEEtran}
\bibliography{bibtex}

\begin{thebibliography}{10}
\providecommand{\url}[1]{#1}
\csname url@samestyle\endcsname
\providecommand{\newblock}{\relax}
\providecommand{\bibinfo}[2]{#2}
\providecommand{\BIBentrySTDinterwordspacing}{\spaceskip=0pt\relax}
\providecommand{\BIBentryALTinterwordstretchfactor}{4}
\providecommand{\BIBentryALTinterwordspacing}{\spaceskip=\fontdimen2\font plus
\BIBentryALTinterwordstretchfactor\fontdimen3\font minus \fontdimen4\font\relax}
\providecommand{\BIBforeignlanguage}[2]{{%
\expandafter\ifx\csname l@#1\endcsname\relax
\typeout{** WARNING: IEEEtran.bst: No hyphenation pattern has been}%
\typeout{** loaded for the language `#1'. Using the pattern for}%
\typeout{** the default language instead.}%
\else
\language=\csname l@#1\endcsname
\fi
#2}}
\providecommand{\BIBdecl}{\relax}
\BIBdecl

\bibitem{review2017}
R.~N. Darmanin and M.~K. Bugeja, ``A review on multi-robot systems categorised by application domain,'' in \emph{2017 25th Mediterranean Conference on Control and Automation (MED)}, 2017, pp. 701--706.

\bibitem{2015survey}
K.-K. Oh, M.-C. Park, and H.-S. Ahn, ``A survey of multi-agent formation control,'' \emph{Automatica}, vol.~53, pp. 424--440, 2015.

\bibitem{quadrotor2023}
J.~Lin, Y.~Wang, Z.~Miao, Q.~Lin, G.~Hu, and R.~Fierro, ``Robust linear-velocity-free formation tracking of multiple quadrotors with unknown disturbances,'' \emph{IEEE Transactions on Control of Network Systems}, vol.~10, no.~4, pp. 1757--1769, 2023.

\bibitem{aditya2019formation}
P.~Aditya, E.~Apriliani, G.~Zhai, and D.~K. Arif, ``Formation control of multi-robot motion systems and state estimation using extended kalman filter,'' in \emph{2019 International Conference on Electrical Engineering and Informatics (ICEEI)}.\hskip 1em plus 0.5em minus 0.4em\relax IEEE, 2019, pp. 99--104.

\bibitem{moorthy2023formation}
S.~Moorthy and Y.~H. Joo, ``Formation control and tracking of mobile robots using distributed estimators and a biologically inspired approach,'' \emph{Journal of Electrical Engineering \& Technology}, vol.~18, no.~3, pp. 2231--2244, 2023.

\bibitem{Bearing2024}
S.~Zhu, K.~Lv, Z.~Yang, C.~Chen, and X.~Guan, ``Bearing-based formation tracking control of nonholonomic mobile agents with a persistently exciting leader,'' \emph{IEEE Transactions on Control of Network Systems}, vol.~11, no.~1, pp. 307--318, 2024.

\bibitem{orientation2023}
J.~Zhao, K.~Zhu, H.~Hu, X.~Yu, X.~Li, and H.~Wang, ``Formation control of networked mobile robots with unknown reference orientation,'' \emph{IEEE/ASME Transactions on Mechatronics}, vol.~28, no.~4, pp. 2200--2212, 2023.

\bibitem{ames2019control}
A.~D. Ames, S.~Coogan, M.~Egerstedt, G.~Notomista, K.~Sreenath, and P.~Tabuada, ``Control barrier functions: Theory and applications,'' in \emph{2019 18th European Control Conference (ECC)}, 2019, pp. 3420--3431.

\bibitem{wang2017safety}
L.~Wang, A.~D. Ames, and M.~Egerstedt, ``Safety barrier certificates for collisions-free multirobot systems,'' \emph{IEEE Transactions on Robotics}, vol.~33, no.~3, pp. 661--674, 2017.

\bibitem{butler2024collaborative}
B.~A. Butler, C.~H. Leung, and P.~E. Par\'{e}, ``Collaborative safety-critical formation control with obstacle avoidance,'' arXiv:2410.03885, 2024.

\bibitem{rai2024safe}
A.~Rai and S.~Mou, ``Safe region multi-agent formation control with velocity tracking,'' \emph{Systems \& Control Letters}, vol. 186, 2024.

\bibitem{li2024safe}
H.~Li, J.~Hu, Q.~Zhou, and B.~K. Ghosh, ``Safe formation control of multiple unmanned aerial vehicles: control design and safety-stability analysis,'' \emph{Control Theory and Technology}, pp. 1--13, 2024.

\bibitem{suganda2024perception}
R.~R. Suganda, T.~Tran, M.~Pan, L.~Fan, Q.~Lin, and B.~Hu, ``Distributed perception aware safe leader follower system via control barrier methods,'' arXiv:2409.11394, 2024.

\bibitem{ballotta2024safe}
L.~Ballotta and R.~Talak, ``Safe distributed control of multi-robot systems with communication delays,'' \emph{IEEE Transactions on Vehicular Technology}, vol.~74, no.~7, pp. 10\,137--10\,150, 2025.

\bibitem{TVT2024}
V.~Bohara and S.~Farzan, ``Adaptive estimation-based safety-critical cruise control of vehicular platoons,'' \emph{IEEE Transactions on Vehicular Technology}, vol.~73, no.~10, pp. 14\,341--14\,353, 2024.

\bibitem{orosz16}
G.~Orosz, ``Connected cruise control: modelling, delay effects, and nonlinear behaviour,'' \emph{Vehicle System Dynamics}, vol.~54, no.~8, pp. 1147--1176, 2016.

\bibitem{swaioa94}
D.~Swaroop, J.~Hedrick, C.~C. Chien, and P.~Ioannou, ``A comparision of spacing and headway control laws for automatically controlled vehicles1,'' \emph{Vehicle System Dynamics}, vol.~23, no.~1, pp. 597--625, 1994.

\bibitem{khalil17}
H.~K. Khalil, \emph{Nonlinear systems}, 2nd~ed.\hskip 1em plus 0.5em minus 0.4em\relax Prentice-Hall, 2011.

\end{thebibliography}

\end{document}